\documentclass[]{aa}
\usepackage{natbib,graphicx,aalongtable}
\def\etal{{\it et al.\ }}

\def\kms{\ {\rm km\,s^{-1}}}

\def\stacksymbols #1#2#3#4{\def\theguybelow{#2}
\def\verticalposition{\lower#3pt}
\def\spacingwithinsymbol{\baselineskip0pt\lineskip#4pt}
\mathrel{\mathpalette\intermediary#1}}
\def\intermediary#1#2{\verticalposition\vbox{\spacingwithinsymbol
\everycr={}\tabskip0pt
\halign{$\mathsurround0pt#1\hfil##\hfil$\crcr#2\crcr 
\theguybelow\crcr}}}

\begin{document}

\title{Enlightening the structure and dynamics of Abell~1942. 
\thanks{based on observations made at the European Southern
Observatory, La Silla (Chile)}}

\author{H.V. Capelato\inst{1} \and  D.~Proust\inst{2} \and G.B.~Lima~Neto\inst{3}
\and W.A. Santos \inst{3} \and L.~Sodr\'e Jr.\inst{3}}

\offprints{D.~Proust} 

\institute{Divis\~ao de Astrof\'{\i}sica, INPE/MCT, 12227-010, 
S\~ao Jos\'e dos Campos/S.P., Brazil
\and 
Observatoire de Paris-Meudon, GEPI, F92195 MEUDON, France 
\and 
Instituto de Astronomia, Geof\'{\i}sica e Ci\^encias Atmosf\'ericas, 
Universidade de S\~ao Paulo (IAG/USP), 05508-090 S\~ao Paulo/SP, Brazil}

\date{Received date; accepted date}

\abstract
{}
{We present a dynamical analysis of the galaxy cluster Abell~1942
based on a set of 128 velocities obtained at the European Southern 
Observatory.}
{Data on individual galaxies are presented and the accuracy of
the determined velocities is discussed as well as some properties of the
cluster. We have also made use of publicly available Chandra X-ray data.}
{We obtained an improved mean 
redshift value  $  z = 0.22513 \pm 0.0008$
and velocity dispersion $\sigma= 908^{+147}_{-139}\kms$. 
Our analysis indicates that inside a radius of $\sim 1.5 h_{70}^{-1}\,$Mpc
($\sim 7\,$arcmin) the cluster is well relaxed, without any remarkable feature
and the X-ray emission traces fairly well the galaxy distribution. Two possible
optical
substructures are seen at $\sim 5\,$arcmin from the centre towards the Northwest
and the Southwest direction, but are not confirmed by the velocity field. These
clumps are however, kinematically bound to the main structure of Abell~1942.
X-ray spectroscopic analysis of Chandra data resulted in a temperature $kT = 5.5
\pm 0.5\,$keV and metal abundance $Z = 0.33 \pm 0.15 Z_{\odot}$. The velocity 
dispersion corresponding to this temperature using the $T_{X}$--$\sigma$ 
scaling relation is in good agreement with the measured galaxies velocities.
Our photometric redshift analysis suggests that the weak lensing signal observed
at the south of the cluster and previously attributed to a ``dark clump'', 
is produced by background sources, possibly distributed as a filamentary 
structure.}
{}


\maketitle

\section{Introduction} \label{Introduction}

In the hierarchical $\Lambda$CDM scenario for structure formation,
clusters of galaxies are the largest coherent and gravitationally bound
structures in the Universe, growing by accretion of nearby galaxy groups 
or even other clusters. These newcomers are often observed as
substructures in the galaxy distribution and, indeed, substructures 
have been detected in a significant fraction of galaxy clusters 
\citep[e.g.,][]{flin06}. Clusters can then be used to trace the cosmological
evolution of structure with time and to constrain cosmological
parameters \citep[e.g.,][]{richstone92,kauffmann93}.

However, clusters comprise a diverse family, presenting a large range of 
structural behaviour and, in order to be useful as cosmological probes, 
the structural and dynamical properties of individual systems should be
determined. Clusters are also complex entities, containing both baryonic 
and non-baryonic matter. In the case of the former, most of the baryons
occupy the cluster volume in the form of a hot gas emitting in X-rays.
Consequently, studies of galaxy clusters aiming to unveil their actual
properties are greatly benefited by multiwavelength observations, in
particular in X-rays for the gas and in the optical for the galaxies and
even the dark matter.

Here we present a study of the cluster of galaxies
Abell~1942. It has richness class 3 and Bautz-Morgan type III.
It is of particular
interest since it has at first glance a quite symmetrical morphology, similar
to Abell~586 which can be used as a laboratory to test different mass
estimators and to analyze its dynamics \citep[as in][]{Cypriano05}.
This cluster was observed in X-ray by several satellites, ASCA, ROSAT, and
Chandra (see Section \ref{x-ray}, below). Finally, A1942 has receive some
attention because a putative mass concentration would have been detected by
its shear effect at 7~arcmin southward from the cluster centre, with no
obvious concentration of bright galaxies at this location \citep{Erben00,
Linden06}: the dark clump. 
A ROSAT-HRI image was also analyzed by the same authors, showing
that the brightest peak of the X-ray emission corresponds with the cluster
centre and its central galaxy. A weak secondary source was also detected at
1~arcmin from the mass concentration. Using ASCA data, \citet{White00} gives 2
temperatures for this cluster: 5.6~keV for a broad band single temperature
fit, and 15.6~keV for a cooling-flow fit, which would corresponds to a
velocity dispersion $\simeq 1800~\kms$ and also a huge cooling flow of 817
$M_{\odot}$/year. 

From the optical data, only 2 velocities of galaxy members
were available, one being the radio-source PKS 1435+038 \citep{Kristian78}.
Moreover, a deep image of the cluster centre \citep{Smail91} shows the
existence of a few lensed arcs, with one being close to the central galaxy. Up
to now, no detailed lens model is available to study the central mass
distribution.

In this paper, we analyze Abell~1942 from its photometric, spectroscopic and
X-ray properties. In Section~\ref{sec:photo}, we give evidence of the
structure and substructures of the cluster from its photometric data.
Section~3 presents the spectroscopic survey of the cluster galaxies in order
to study the velocity dispersion in the cluster centre, as well as its
variations with the radius until the measured limit of the shear up to
8~arcmin (equivalent to a radius of $1.7 h^{-1}_{70}$~Mpc at the cluster
redshift). In Section~4 we analyze the X-ray data. The velocity analysis is
detailed in Section~5. With such a set of velocities we build in Section~6 a
detailed image of the cluster dynamics and mass distribution. Moreover we
analyze the velocity distribution of the galaxies located close to the mass
concentration area in order to know its nature as distant cluster or
concentration of matter associated with the main cluster. We adopt here,
whenever necessary, $H_{0}= 70\, h_{70}\kms$Mpc$^{-1}$, $\Omega_{M} = 0.3$ and
$\Omega_{\Lambda} = 0.7$.

\section{Abell 1942 photometric data}\label{sec:photo}

\begin{figure}[htb]
\centering \includegraphics[width=\columnwidth]{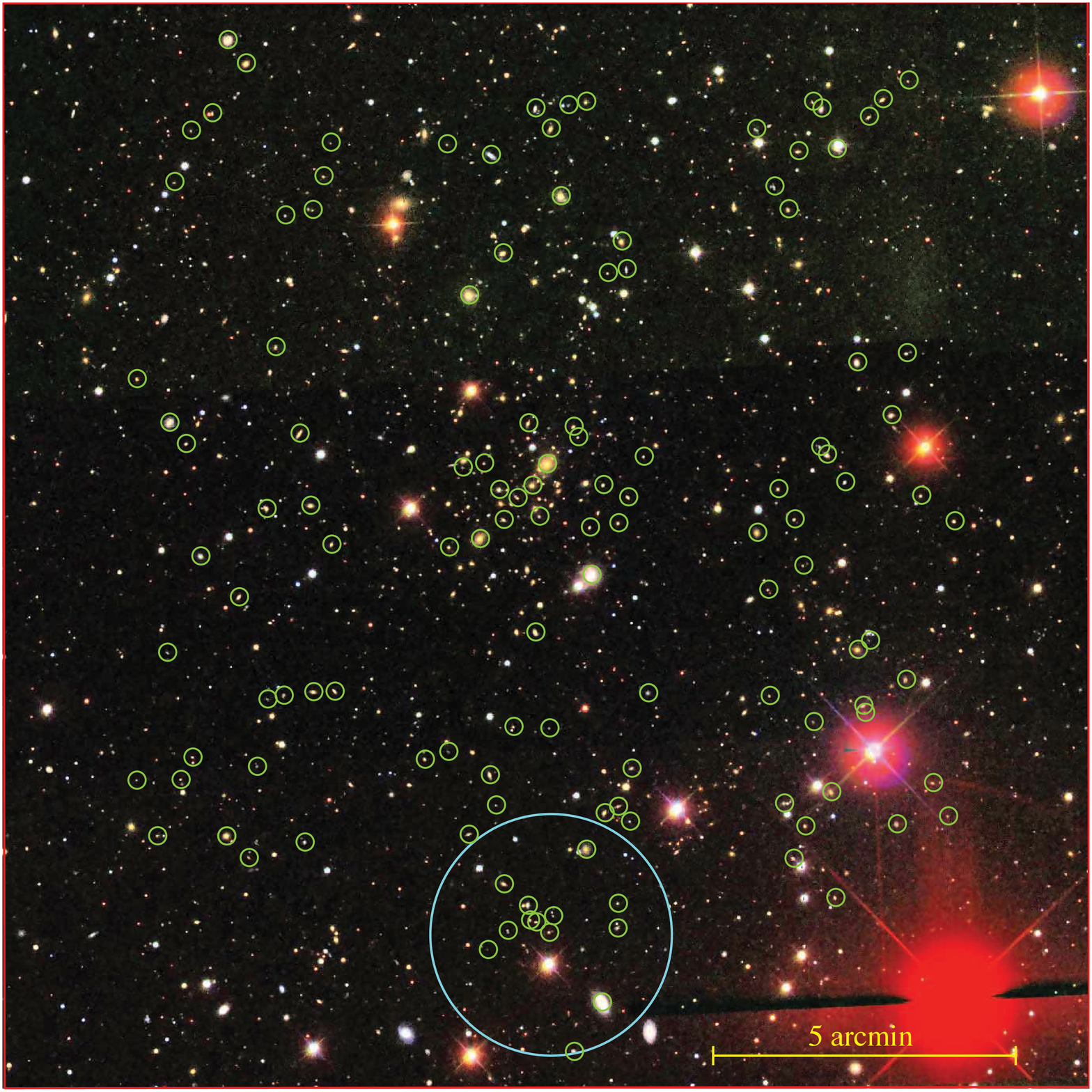}
\caption[]{A $18  \times 18$  arcmin true colour  image from  the SDSS
centered  at the  cluster  centre. The  larger  circle, with  2~arcmin
radius,  corresponds to  the  dark matter  clump  region suggested  by
\citet{Erben00}.    Small  circles   superimposed  to   galaxy  images
correspond to the spectroscopic observations reported in this work}
\label{fig:map}
\end{figure}

The Abell~1942 cluster of galaxies is within the area observed by the Sloan
Digital Sky Survey (SDSS) \footnote{http://www.sdss.org/; funding for the 
SDSS and SDSS-II has been
provided by the Alfred P. Sloan Foundation.}. In this
paper we adopt the photometric data from its Data Release Six. Fig.
\ref{fig:map} shows a $18 \times 18$ arcmin square SDSS image centered at the
cluster centre [assumed to be at the position of the brightest cluster galaxy,
$m_r=16.23$, at at R.A.~$14^{\rm h}\,38^{\rm m}\,21.9^{\rm s}$,
Dec~$+03^{o}~40'~13''$(J2000)] and extending to the South to show the region
where a dark mass concentration would have been detected \citep{Erben00}.

\begin{figure}[!htb]
\centering \includegraphics[width=\columnwidth]{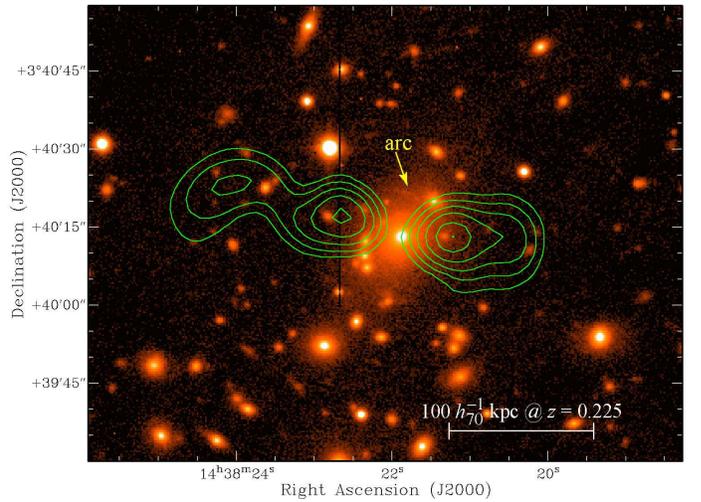}
\caption[]{A $V$-band image of the central part of the A1942 cluster taken
with EFOSC at the 3.60m ESO telescope. The gravitational arc is
indicated by the arrow. The green contours are isophotes of the 21cm
VLA-FIRST, logarithmically spaced. The absence of a head-tail radio
morphology indicates that the brightest cluster galaxy is at rest with respect
to the intracluster gas.}
\label{fig:centro}
\end{figure}

\citet{Smail91} announced  the presence of a lensed  arc candidate in
the  cluster centre  from  $V$ CCD  frames  taken on  the Danish  1.5m
telescope at La  Silla. During our observations, we  also obtained $V$
images which  show the centre resolved in  independent components with
the  lensed  arc clearly  visible.   This  image  is shown  in  Fig.
\ref{fig:centro},  to  which we  superimposed  the  21cm  emission
isophotes from the VLA-FIRST (Faint Images of the Radio Sky at 
Twenty-centimeters) survey \citep{Becker94}.

Within  the  region  shown  in  Fig. \ref{fig:map},  there  are  674
galaxies with dereddened $r$~magnitudes  between 15.36 and 21.0, 17 of
them with spectroscopic redshifts in the SDSS database. To this sample
we added  128 new  spectroscopic redshifts, 9  of them in  common with
SDSS. About 11 of these new  redshifts are in a 200~arcsec side square
region centered on the putative dark clump claimed by \citet{Erben00}.

\begin{figure}[htb]
\centering
\includegraphics[width=\columnwidth]{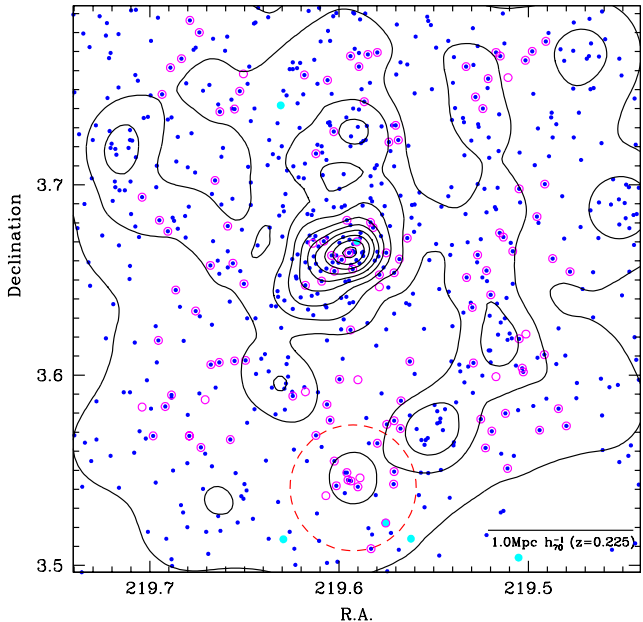}
\caption[]{The projected density map of the galaxies brighter than $m_r= 21.0$
in the 18~arcmin field of Fig. \ref{fig:map}. Filled circles denote SDSS
galaxies with $m_r \leq 21.0$, with larger (cyan) circles for the brightest
ones: $14.19 \leq m_r \leq m_{r_1}$, with $m_{r_1} = 16.23$. Open circles
denotes galaxies with measured spectroscopic redshifts. Notice that some of
these galaxies are fainter than the $21^{\rm mag}$ limit (see Table 1). The
dashed circle encloses the region of the alleged dark clump of
\citet{Erben00}. Contours are equally spaced by $\Delta_{\Sigma} = 0.4$ with
highest levels at $\Sigma_{max} = 3.81$, with $\Sigma$ in units of $ 10^{-3}
\rm{gal~arcsec^{-2}}$. }
\label{fig:iso}
\end{figure}

Fig. \ref{fig:iso} shows the contour map of the projected density of the
674~galaxies brighter than $m_r= 21.0$. We adopt this magnitude limit because
of our photometric redshifts are reliable up to this value (see Section
\ref{DC}). The centre of the cluster appears clearly at $R.A. =
219.596^{\circ}, Dec = +3.663^{\circ},~ \sim 30 arcsec$ SE from the
dominant cluster galaxy $(219.591^{\circ},+3.670^{\circ})$. The
isodensity contours seems slightly elongated in the NW-SE direction, with no
significant substructures except perhaps to the north of the cluster centre.
At SW ($\simeq 219.54^{\circ}$, $+3.63^{\circ}$) we find a not very
significant overdensity which could be identified to the ``C'' component
noticed by \citet{Erben00} in their figure~10. A small concentration of
galaxies is also seen in the region of the alleged dark clump (dashed circle
in Fig. \ref{fig:iso}). We will return to this point in Section \ref{DC}.

\section{Spectroscopic observations and Data Reductions}\label{sec:ObservationsReduction}

The A1942 spectra used in this paper have been obtained with the 3.60m
telescope at  ESO-La Silla (Chile) in  two runs during a  total of ten
half-nights from 5 to 10 June 2005  and 30 April to 3 May 2006 (with 4
cloudy  nights). A  $14.5 \times  14.5$ arcmin  field centered  at the
brightest cluster galaxy was tiled into 9~($3 \times 3$) adjacent $5.4
\times 5.4$~arcmin fields with a 34 arcsec overlap between them. A 10th
field towards the South was added, corresponding to the location of the
dark clump  (these fields are shown  in Fig. \ref{fig:mapa_grupos}
below). The instrumentation used was the ESO Faint Object Spectrograph
and  Camera  (EFOSC)   with  the  grism  8  giving   a  dispersion  of
0.99{\AA}/pix.

The targets  were preferentially  selected from a  SDSS sample  of 327
galaxies brighter than (undereddened)  $m_r= 20.5$, from which only 97
were picked out given the constraints  of the FOV's of EFOSC. In order
to fill the spectroscopic masks with punched slitlets, we added to our
list 98 galaxies candidates (as classified by SDSS), fainter than this
limit,  totalizing 195  targets.   Standard stars  were also  observed
during each night for flux calibration of the spectra.
 
The  data  reduction  was   carried  out  with  IRAF\footnote{IRAF  is
distributed by the National Optical Astronomy Observatories, which are
operated by the Association of Universities for Research in Astronomy,
Inc.,   under  cooperative   agreement  with   the   National  Science
Foundation.} using the MULTIRED  package.  Radial velocities have been
determined  using   the  cross-correlation  technique  \citep{Tonry79}
implemented in  the RVSAO package \citep{Kurtz91,  Mink95} with radial
velocity standards  obtained from observations of  late-type stars and
previously well-studied galaxies.

About 141  of the 195  observed spectra had  S/N high enough  to allow
redshift estimates. About  14 of them were found to  be stars. We were
thus left with a total of 128 galaxies with measured redshifts, one of
which being detected  as a possible binary system  (galaxies \#113 and
\#116  in Table~A.1).  About  9  of these  spectra are  in common  with
SDSS. About 117 of these are  situated in the central square region of
14.5 arcmin side, whereas the remaining 11 are located in a 5.4 arcmin
side square field  centered on the on the  putative dark clump claimed
by \citet{Erben00}.

Table~A.1\footnote{Table~A.1 is also available in electronic form at the CDS
via anonymous ftp 130.79.128.5} lists positions, dereddened magnitudes $u$,
$g$, $r$, $i$ and $z$ (SDSS database), photometric redshifts and errors (see
Section \ref{DC} and the Appendix) and the heliocentric spectroscopic
redshifts from the present work. Redshifts errors were derived following
\citet{Tonry79}. The values of their R statistics (defined as the ratio of the
correlation peak height to the amplitude of the antisymmetric noise) are
listed in the $notes$ column.

\subsection{The Completeness of the Spectroscopic Sample}\label{completeness}

\begin{figure}[htb]
\centering
\includegraphics[width=\columnwidth,angle=0]{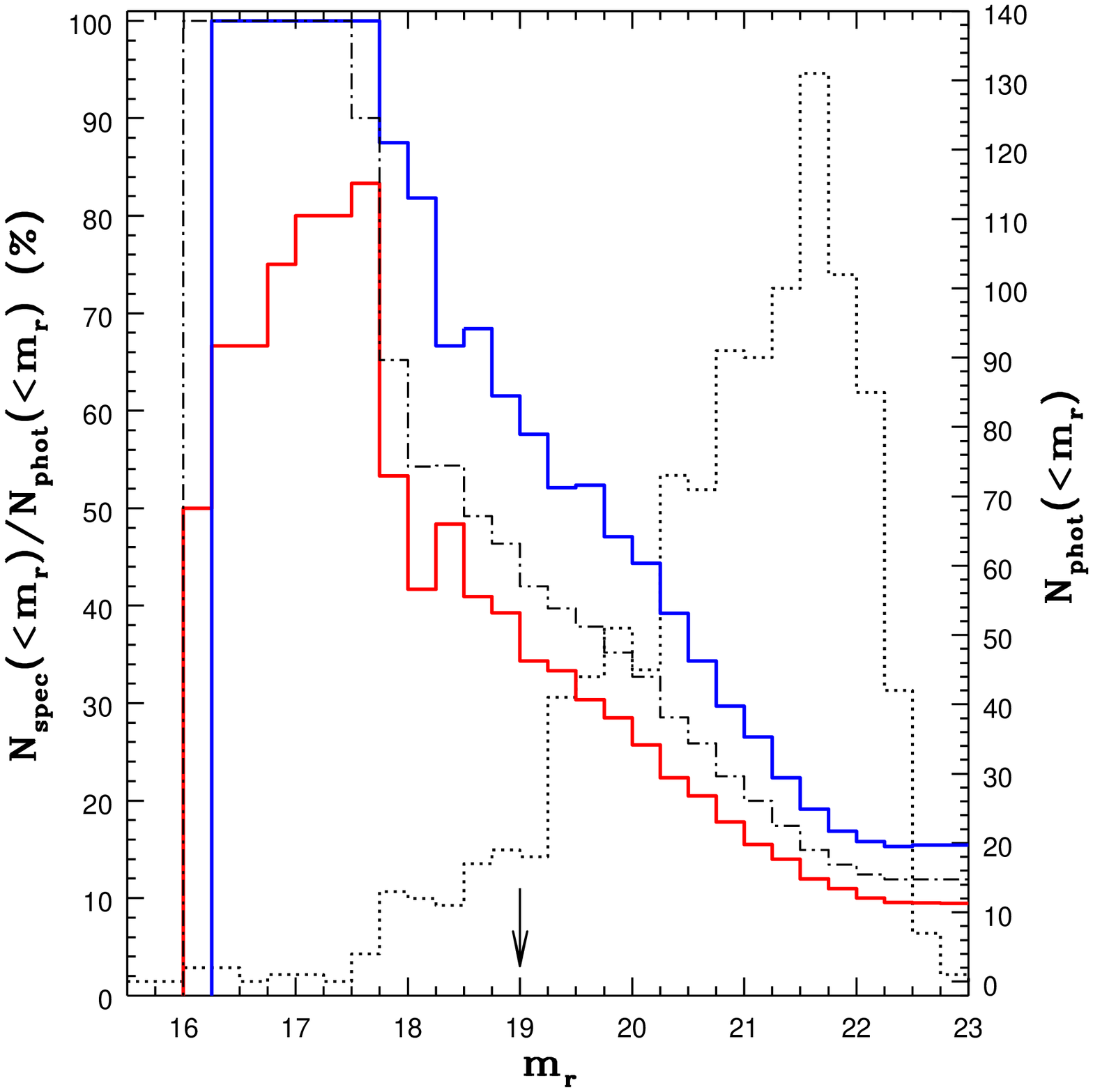}
\caption[]{The completeness of the spectroscopic sample as a function
of magnitude.   \textsf{Dot-dashed lines}: square  region 14.5 arcmin
side centered  in the cluster. \textsf{Red  continuous lines}: central
circular region  of radius 350~arcsec. \textsf{Blue continuous
line}: its complementary region. \textsf{Dotted lines}: the luminosity
distribution of  galaxies of the photometric sample,  with values been
read in  the right  vertical axis.  The  arrow points at  the predicted
value of $M^*$ for an Abell cluster at this redshift.}
\label{fig:compl}
\end{figure}

We  estimated  the  completeness  of  the spectroscopic  sample  as  a
function of  the magnitude  as the ratio  of cumulative counts  of the
spectroscopic   samples    to   that   of    the   photometric   ones,
$N_{spec}(<m_r)/N_{phot}(<m_r)$,   all  computed  within   the  square
14.5~arcmin  side region  containing most  of our  spectral  data. The
results  are displayed  in Fig.  \ref{fig:compl}, together  with the
magnitude distribution of the photometric sample. The completeness was
computed  both for the  entire 14.5  $\times$ 14.5~arcmin  region over
which the photometric sample was defined (light continuous lines), and
over a  central circular region  of 350~arcsec radius  (red continuous
lines)  which  roughly  corresponds  to  the region  where  the  X-ray
emission is  detected (Section \ref{x-ray}). Following  the results of
\citet{Paolillo01}, the mean $M^*$-magnitude of rich Abell clusters is
$M^*_r = -21.24$ (after  applying a $0.2^{\rm mag}$ correction between
his DPOSS magnitudes and the  SDSS magnitudes used here) which, at the
redshift of A1942 (see  Section \ref{Velocity analysis}), $z = 0.225$,
gives  $m^*_r  =  19.00$,  indicated   by  an arrow  line  in  Fig.
\ref{fig:compl}.

From  this plot  it can  be seen  that the  spectroscopic  data, which
samples nearly uniformly the  entire 14.5 $\times$ 14.5 arcmin region,
is highly incomplete  for magnitudes fainter than $m_r \sim
18$. Of course  this is due to the clustering of  galaxies on the very
centre of the region, as it may readily seen in Fig. \ref{fig:compl}
by comparing the completeness  of the central circular region defining
the virialized cluster  ($\sim 40\%$ at $m_r \sim  m^*_r$), to that of
its  complementary  region ($\sim  63\%$  at  $m_r  \sim m^*_r$;  blue
continuous lines).

\section{X-ray data}\label{x-ray}

Abell 1942 was first observed by the Einstein satellite Image Proportional
Counter (IPC) in 1979 and 1980. Then, it was again observed both by the ROSAT
High Resolution Imager (HRI) and ASCA in 1995. In 2003, a 58.3~ks exposure was
acquired with Chandra Advanced CCD Imaging Spectrometer-Imager (ACIS-I)
detector (ObsID 3290, PI. G.~P. Garmire). In 2007 Abell 1942 was again
observed by Chandra but only for 5.18~ks. We have downloaded only the ObsID
3290 observation from Chandra X-ray Centre (CXC) archives in order to analyze
it.

The morphology of the X-ray emission observed by  Chandra was already
analyzed by \citet{Linden06}, paying  close attention to the region of
the putative dark clump. Contrary to \citet{Erben00} analysis based on
ROSAT data, they do not detect any significant extended X-ray emission
at the supposed location of the dark clump with Chandra data.

Here, we concentrate on the X-ray analysis of the cluster itself which
was so far neglected. The data,  taken in Very Faint mode, were reduced
using CIAO version 3.4\footnote{\texttt{http://asc.harvard.edu/ciao/}}
following the  Standard Data Processing,  producing new level 1  and 2
event files. The level 2 event file was further filtered, keeping only
events with grades%
\footnote{The  grade of  an  event  is a  code  that identifies  which
pixels, within  the three pixel-by-three pixel island  centered on the
local  charge maximum,  are above  certain amplitude  thresholds.  The
so-called \textit{ASCA}  grades, in the  absence of pileup,  appear to
optimize           the           signal-to-background           ratio.
\texttt{http://cxc.harvard.edu/}} 0, 2, 3, 4 and 6. We checked that no
afterglow  was  present and  applied  the  Good  Time Intervals  (GTI)
supplied  by the  pipeline.   Only some  mild  background flares  were
observed and the corresponding  time intervals were filtered away. The
final total livetime is 54.66~ks.

We   have  used   the  CTI-corrected   ACIS  background   event  files
(``blank-sky''),     produced      by     the     ACIS     calibration
team\footnote{\texttt{http://cxc.harvard.edu/cal/Acis/WWWacis\_cal.html}},
available  from  the calibration  data  base  (CALDB). The  background
events were  filtered, keeping the  same grades as the  source events,
and then were  reprojected to match the sky  coordinates of Abell 1942
ACIS-I observation.

We restricted  our analysis to  the range [0.3--7.0~keV],  since above
$\sim  7.0$~keV, the  X-ray observation  is largely  dominated  by the
particle background.

\begin{figure*}[!htb]
  \centering \includegraphics[width=18cm]{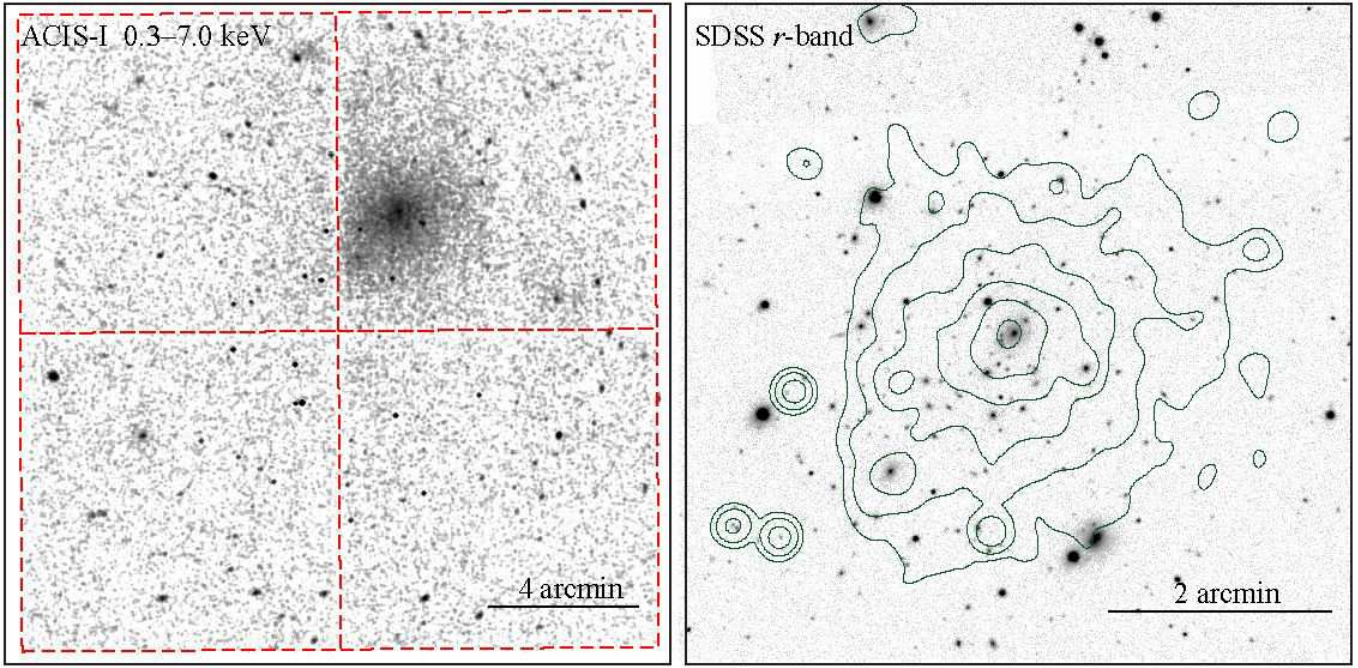}
  \caption[]{Left: Chandra ACIS-I image of the cluster. Also shown for
  reference is  the position of ACIS-I four CCDs (red  dashed lines).
  Right:  SDSS  $r$-band  image  of  the  central  region  with  X-ray
  isophotes contours superposed (green solid lines).}
  \label{fig:A1942_ACIS_DSSr}
\end{figure*}

\subsection{Spectral analysis}

For   the   spectral  analysis,   we   have   computed  the   weighted
redistribution  and ancillary  files  (RMF and  ARF)  using the  tasks
\textsc{mkrmf} and  \textsc{mkwarf} from  CIAO. These tasks  take into
account the extended nature  of the X-ray emission. Background spectra
were constructed from the blank-sky  event files and were extracted at
the same regions (in detector  coordinates) as the source spectra that
we want to fit.

The spectral fits were done using \textsc{xspec}~v11.3.2. The X-ray spectrum
of each extraction region was modelled as being produced by a single
temperature plasma and we employed the \textsc{mekal} model \citep{Kaastra93,
Liedahl95}. The photoelectric absorption -- mainly due to neutral hydrogen --
was computed using the cross-sections given by \citet{Balucinska92}, available
in \textsc{xspec}. We have used metal abundances (metallicities) scaled 
to \citet{Anders89} solar values.

The  overall  spectrum  was  extracted  within a  circular  region  of
78~arcsec ($280  h_{70}^{-1}$~kpc at $z=0.225$) centered  on the X-ray
emission peak. It was re-binned with the \textsc{grppha} task, so that
there are  at least 10 counts  per energy bin. This  radius was chosen
because most of  the cluster emission is in this  region, we can avoid
the CCDs  gaps, and  we have  all the spectrum  extracted in  a single
ACIS-I  CCD.   The  Chandra   ACIS-I  image  is  displayed  on  Fig.
\ref{fig:A1942_ACIS_DSSr} with the $r$-band image of the same region.

\begin{figure}[!htb]
    \centering
    \includegraphics[width=\columnwidth]{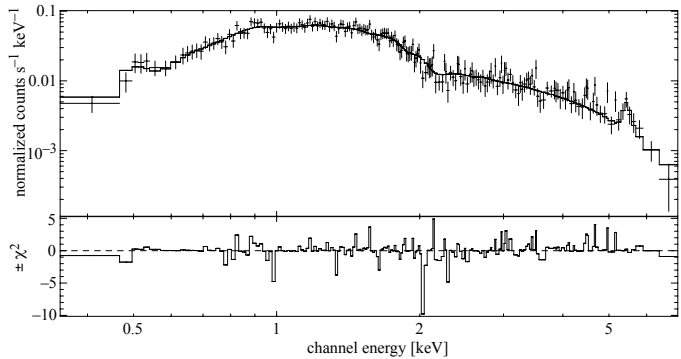}
    \caption[]{\textit{Chandra} ACIS-I X-ray spectrum extracted in the
    central  78~arcsec  ($280  h_{70}^{-1}\,$kpc)  superposed  with  a
    \textsc{mekal} plasma spectrum. \textsl{Top:} fit with $N_{\rm H}$
    free. \textsl{Bottom:}  the residuals  are shown expressed  as the
    $\chi^{2}$ contribution of each binned energy channel.
    \label{fig:a1942Mekalcirc}}
\end{figure}

Table~\ref{tbl:mekalfits} summarizes the spectral fitting results. The
best fit temperature, $kT$, varies between 5.3 and 5.6~keV depending on the
free and fixed parameters, with an error bar of about 0.4~keV at $1\sigma$
confidence level. The metallicity, independently of parameters kept fixed, is
$0.33 Z_{\odot}$, with an error bar $0.09 Z_{\odot}$ at $1\sigma$
confidence level. When we left the hydrogen column density free, we
obtain $N_{\rm H} = (3.9 \pm 1.2) \times 10^{20}$~cm$^{-2}$ which agrees, within
less than $2 \sigma$ error bars, with the Galactic value, $N_{\rm H} = (2.61
\pm 0.07) \times 10^{20}$~cm$^{-2}$ given by the Leiden-Argentina-Bonn
Survey \citep{Kalberla05}.

\begin{table}[!htb]
 \centering
 \caption[]{Results of the X-ray spectral fits in a circular region of
 radius 78~arcsec around  the centre. ``dof'' is short  for degrees of
 freedom.}
 \label{tbl:mekalfits}
  \tabcolsep=0.5\tabcolsep
 \begin{tabular}{ccccc}
 \hline\\[-2ex]  
 $N_{\rm  H} [10^{20}$cm$^{-2}$]  &  $kT$  [keV] &  $Z
 [Z_{\odot}]$  &  redshift  &  $\chi^{2}/$dof  \\[3pt]  
 \hline\\[-2ex]
 $3.9 \pm 1.2$  &  $5.3 \pm 0.4$  & $0.33 \pm 0.09$ &  0.225             & 196.2/231 \\[3pt]
 2.61           &  $5.6 \pm 0.3$  & $0.33 \pm 0.09$ &  0.225             & 196.7/232 \\[3pt]
 $3.9 \pm 1.2$  &  $5.3 \pm 0.4$  & $0.33 \pm 0.09$ & $0.222  \pm 0.007$ & 196.3/230 \\ 
 \\[-2ex] 
 \hline
 \end{tabular}
 \begin{flushleft}
     Note: Error bars are $1 \sigma$ confidence level.  Values without error
     bars are kept fixed.
 \end{flushleft}
\end{table}

We  have also  used the  \textsc{vmekal} model,  where  the individual
metal abundances  are fitted independently. The  results are presented
in Table~\ref{tab:vmekal}.

\begin{table}[!htb]
 \centering
 \caption[]{Results  of   X-ray  spectral  fits with  varying  metal
 abundances.    The    error   bars   are    $1   \sigma$   confidence
 level. Temperature is in keV and the abundances in solar units.}
 \label{tab:vmekal}
\tabcolsep=0.2\tabcolsep
\begin{tabular}{cccccc}
    \hline
    O   &  Mg   &  Si  &   Ca  &   Fe  &   Ni   \\
    \hline\\[-2.5ex]
  $0.9_{-0.9}^{+1.1}$  &  $1.9 \pm 1.1$ & $0.5 \pm 0.5$  & 
$2.2 \pm 2.1$ &  $0.36 \pm 0.09$ & $2.7 \pm 2.5$  \\[0.5ex]
\hline
\hline\\[-2.5ex]
    \multicolumn{2}{l}{$kT = 5.4 \pm 0.3$ keV} & &
    \multicolumn{2}{l}{$\displaystyle{\frac{\chi^{2}}{\rm   dof} = \frac{192.3}{227}}$}\\[1.5ex]
    \hline
\end{tabular}
\end{table}

There is a marginal evidence of an over-abundance of $\alpha$-elements, mainly
Mg (considering the error bars). The best fit value for Ni is also very high,
7.5 times the Fe abundance but, again, error bars prevent us from discussing
further this question. We only present the individual metal abundances for
completeness.

\subsection{Radial profiles}

\subsubsection{Temperature profile}\label{sec:ktfit}

\begin{figure}[htb]
    \centering
    \includegraphics[width=\columnwidth]{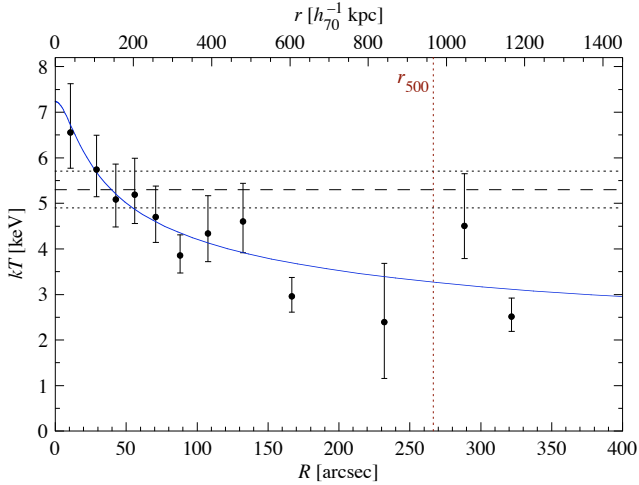} 
   \caption[]{Temperature profile. The error bars are $1 \sigma$ 
   confidence
   level. The horizontal dashed line is the mean temperature inside 78~arcsec,
   $kT = 5.3 \pm 0.4$ keV, the horizontal dotted lines correspond to $1\sigma$
   confidence level of the mean temperature. The continuous curve is the
   best-fit polytrope temperature profile. The vertical line shows $r_{500}$.}
    \label{fig:kT_perfilNHfix}
\end{figure}

The radial temperature profile was obtained in concentric circular annuli
where, for each annulus, a spectrum was extracted and fitted following the
method described above, except that the hydrogen column density was kept fixed
at the mean best-fit value found inside 78~arcsec (i.e., $N_{\rm H} = 3.9
\times 10^{20}$cm$^{-2}$). The annuli are defined by approximately the same
number of counts (1000 counts, background corrected), so that the
signal-to-noise in each annulus spectrum is roughly constant.
Fig.~\ref{fig:kT_perfilNHfix} shows the temperature profile.

The temperature profile shown in Fig.~\ref{fig:kT_perfilNHfix} presents a
visible gradient outwards. We therefore used a simple analytical temperature
profile, described by a polytropic equation of state, to fit the observed data
points. Although it is not clear if the ICM gas temperature is well described 
by a polytropic equation of state, some observations suggest that a polytrope 
with index $1.1 \ga \gamma \la 1.3$ may be empirically used to describe the
temperature radial profile \citep[e.g.][]{Markevitch99,LimaNeto03,Cypriano05}.

The use of a polytropic temperature profile has also the benefit of being
easily deprojected. Thus, we have fitted a temperature profile given by:
\begin{equation}
    T(r) = T_{0} \left[ 1 + \left(\frac{r}{r_{c}}\right)^{2} 
    \right]^{-3\beta (\gamma-1)/2} \, ,
    \label{eq:TempProfile}
\end{equation}
which is the temperature of a polytropic gas that follows a
$\beta$-model radial profile. Here, $r_{c}$ and $\beta$ are the values
obtained with the $\beta$-model fitting of the brightness surface profile (see
Sect.~\ref{sec:xrayProfil} below), and $T_{0}$ is the central temperature.
Notice that only $\gamma$ and $T_{0}$ are free parameters.

A standard least-square fit of Eq.~(\ref{eq:TempProfile}) results in $T_{0} =
7.2 \pm 0.6$ keV and $\gamma = 1.23 \pm 0.04$, with a reduced $\chi^{2} =
1.1$; the best-fit polytropic temperature profile is ploted in
Fig.~\ref{fig:kT_perfilNHfix}. The fitted polytropic index is below $5/3$, the
value of an ideal gas, suggesting that the gas may be indeed in adiabatic
equilibrium \citep[see, eg.,][Sect.~5.2]{Sarazin88}.

\subsubsection{X-ray brightness profile}\label{sec:xrayProfil}

\begin{figure}[!htb]
  \centering
  \includegraphics[width=\columnwidth]{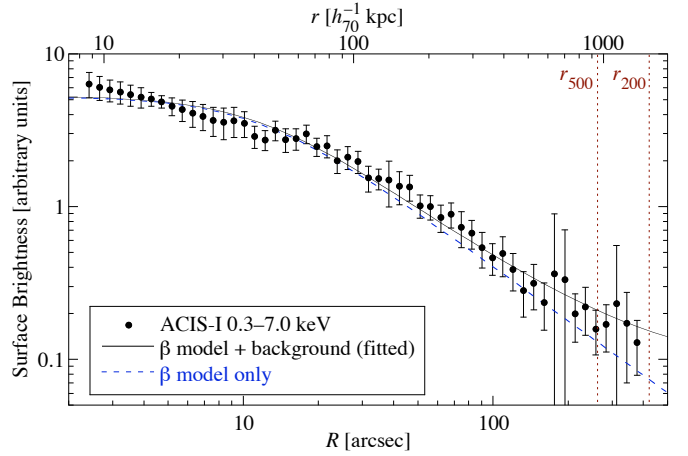}
  \caption[]{X-ray brightness profile. The  full line shows the actual
  fit of  a $\beta$-model  plus a constant  background, with  $\beta =
  0.38\pm 0.01$ and $rc = 12.5 \pm 1.6\,$arcsec. The dashed line shows
  only  the  cluster  contribution  to the  brightness  profile.   The
  vertical lines show $r_{200}$ and $r_{500}$ (see below).}
  \label{fig:A1942_SBBetaFits}
\end{figure}

The X-ray brightness profile of  Abell 1942 was obtained with the task
\textsc{ellipse} from STSDAS/IRAF.
The image we have used was in the [0.3--7.0 keV], corrected by the exposure
map and binned so that each image pixel has 2~arcsec. Prior to the ellipse
fitting, the CCDs gaps and source points were masked. The brightness profile,
shown in Fig.~\ref{fig:A1942_SBBetaFits}, could be measured up to $\sim
350$~arcsec ($1.26\, h_{70}^{-1}$Mpc) from the cluster centre.

In order to describe the surface brightness radial profile, we use the
$\beta$-model \citep{Cavaliere76}:
\begin{equation}
     \Sigma_{X}(R)       =      \Sigma_{0}       \left[       1      +
\left(\frac{R}{R_{c}}\right)^{2}\right]^{-3\beta      +     1/2}     +
\mbox{background} \, ,
\end{equation}
where we have  added a constant to take  into account the contribution
of the background (both cosmic and particle).  A least-squares fitting
gives $\beta =  0.38 \pm 0.01$ and $R_{c} =  12.5 \pm 1.6$~arcsec ($45
\pm 6\, h_{70}^{-1}$kpc).  If we  assume that the gas is approximately
isothermal and distributed with  spherical symmetry, there is a simple
relation between  the brightness profile  and the gas  number density,
$n(r)$, i.e.,
\begin{equation}
 n(r) = n_{0} \left[ 1 +
           \left(\frac{r}{r_{c}}\right)^{2}\right]^{-3\beta/2} \, ,
     \label{eq:gasDens}
\end{equation}
where  $R_{c} =  r_{c}$ (capital  indicates projected  2D coordinates,
lower case indicates 3D coordinates).

In order to estimate the central density, $n_{0}$, which is related to
$\Sigma_{0}$, we integrate the bremsstrahlung emissivity along the
line-of-sight in the central region. The result was compared with the flux
obtained by spectral fitting of the same region, the normalization parameter
of the thermal spectral model in \textsc{xspec}. This parameter, in
turn, is proportional to $n_{e} n_{\rm H} \approx 1.2\, n_{\rm H}^{2}(r)$
(where $n_{e}$ and $n_{\rm H}$ are the electron and proton number densities,
respectively). We obtain thus $n_{0} = (8.8 \pm 0.9) \times
10^{-3}\,$cm$^{-3}$ (we drop the index H hereafter).

Abell  1942 presents a  rather steep  surface brightness  profile (see
Fig.  \ref{fig:A1942_SBBetaFits}).    Such  a  profile   is  usually
associated  with  a  relaxed  cluster  with a  cool-core,  a  drop  in
temperature of a factor $\sim 3$ in the centre compared to the maximum
temperature.

However, we note that this cluster does not present any sign of a cool-core in
the central part, at $r \approx 100\, h_{70}^{-1}\,$kpc, the smallest radius
where we can extract a meaningful spectrum and measure the temperature. We
actually measure an increase of the temperature from $r \approx 300\,
h_{70}^{-1}\,$kpc towards the centre. We may be failing to detect a cool-core
either because we lack the necessary spatial resolution or the intracluster
gas is not cooling due to some physical heating process -- as is the
case of numerous clusters \citep[e.g.,][]{Arnaud05,Snowden08}.

Heating  by cluster  merging may  be  a possible  mechanism. There  is
indeed  a substructure at  1.7~arcmin towards  the southeast  from the
cluster centre (see Fig.\ref{fig:A1942_ACIS_DSSr}). However, we may be
simply not detecting  an eventual drop in temperature  because we lack
the resolution. Using a  sample of 20~clusters, \citet{Kaastra04} show
that the radius  ($r_{c}$ in their paper) where  the temperature drops
in  cooling-flow clusters is,  with 2  exceptions, smaller  than $70\,
h_{70}^{-1}\,$kpc.

\subsubsection{Mass profile}

\begin{figure}[htb]
    \centering
    \includegraphics[width=\columnwidth]{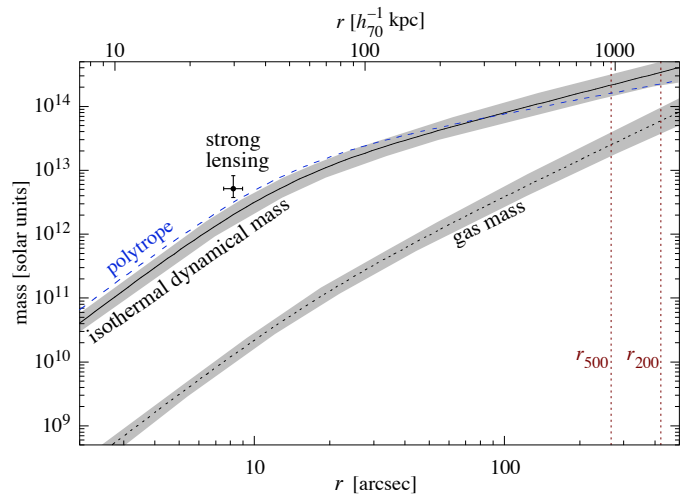}
\caption[]{X-ray dynamical mass profiles and gas mass profile (dotted line).
The full line corresponds to the dynamical mass computed assuming an
isothermal $\beta$-model; the dashed line is corresponds to the polytropic
temperature profile. The vertical lines correspond to $r_{200}$ and $r_{500}$.
The grey shaded areas correspond to $1\sigma$ confidence level error bars. The
single point named ``strong lensing'' is our estimate based on the position of
the arc seen in Fig.~\ref{fig:centro} (see text for details).}
    \label{fig:massPerfil_BetaIsoA1942}
\end{figure}

We compute  the gas  mass simply by  integrating the density  given by
Eq.~(\ref{eq:gasDens}),  assuming spherical  symmetry  which, in  this
case, seems a  good approximation when we exclude  the substructure at
the southeast  (see Fig.~\ref{fig:A1942_ACIS_DSSr}). The  growth gas
mass profile is shown in Fig.~\ref{fig:massPerfil_BetaIsoA1942}.

The total mass (``X-ray dynamical mass'') is estimated assuming either
an isothermal temperature equal to the emission weighted mean temperature of
the cluster, or the deprojected polytropic temperature radial profile fitted in
Sect.~\ref{sec:ktfit}. Assuming hydrostatic equilibrium and spherical
symmetry, the corresponding dynamical mass for the $\beta$-model is
shown in Fig.~\ref{fig:massPerfil_BetaIsoA1942}. The values for the
$r_{200}$ and $r_{500}$ radii derived from this model are, respectively,
$1.53\,h_{70}^{-1}\,$Mpc and $0.97\,h_{70}^{-1}\,$Mpc.

The difference in the dynamical mass estimates using the isothermal and
polytropic temperature profiles are quite small. For the polytropic model, the
mass profile rises more steeply near the center and than has a slower growth
beyond $r \sim 300 h_{70}^{-1}\,$kpc.

We have also estimated the total mass using the gravitational arc shown in
Fig.~\ref{fig:centro}, discovered by \citet{Smail91} assuming that
this strong arc is located at the Einstein radius, $R_{E}$. Measuring the
distance from the centre of the dominant galaxy, we have $R_{E} = 8.2 \pm
0.8$~arcsec or, at $z = 0.225$, $R_{E} = (29.6 \pm 2.9) h_{70}^{-1}\,$kpc. We
then suppose, for simplicity, that the dynamical mass may be modelled by a
singular isothermal sphere. If the background lensed galaxy is located between
$0.6 < z < 4.0$ then the total mass inside $R_{E}$ is within $(3.6 < M < 8.0)
\times 10^{12}\,M_{\odot}$. We show this mass estimate in
Fig.~\ref{fig:massPerfil_BetaIsoA1942}, in comparison with the dynamical mass
derived with X-ray data.

The gravitational  lensing mass  is about a  factor 2 larger  than the
mass obtained with X-ray data,  but the difference is only significant
at  less than  $2  \sigma$  confidence level.   Indeed,  this kind  of
situation, where the ``lensing mass'' is larger than ``X-ray mass'' is
known for some time \citep[e.g.]{Cypriano04}. While the dynamical mass
estimated  by  lensing effects  may  overestimate  the  total mass  by
including  all contributing  mass along  the line-of-view,  X-ray data
derived masses  may sometimes under-estimate the total  mass, as shown
by numerical simulations by \citet{Rasia06}.

\section{Analysis of the Velocity Distribution}\label{Velocity analysis}

\subsection{Kinematical Structures}\label{kinematical structures}

\begin{figure}[htb]
\centering
\includegraphics[width=\columnwidth]{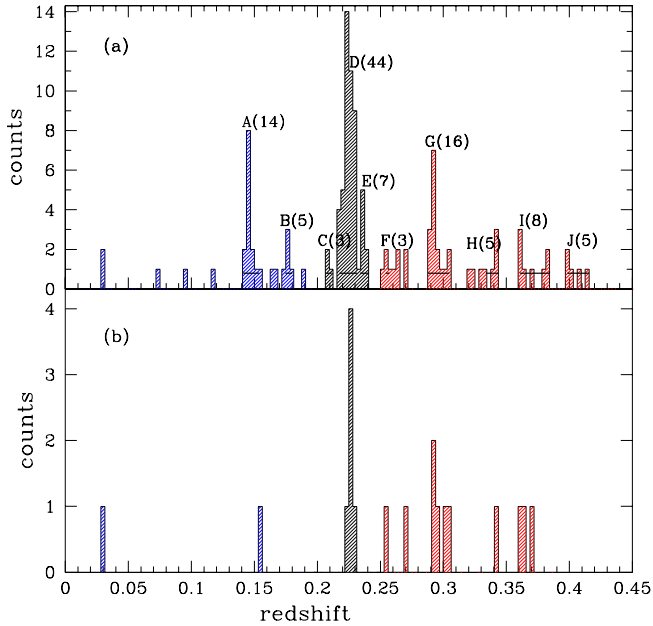}
\caption[]{{\sl Upper panel:} Histogram of the redshift distribution of 
our spectroscopic sample. Letters identify the kinematical structures
found with the gap analysis  The number  of identified  members are  in
parentheses. The horizontal lines inside the histograms of groups give 
the range of the groups as determined from the gap analysis. 
{\sl Lower panel:} redshift distribution for the circular region
enclosing the DC region.}
\label{fig:histoz}
\end{figure}

We used  the ROSTAT routines  \citep{Beers90} to analyze  the velocity
distribution  of  the  spectroscopic  sample  given  in  Table~A.1. We
identified the kinematical structures using the method of the weighted
gap analysis,  as discussed in  \citet{Ribeiro98}.  A weighted  gap is
defined by  $ y_i = (w_ig_i)^{-1/2} \, ,$
where  the $g_i{\rm 's}$ are  the measured  gaps between  the ordered
velocities, and the $w_i{\rm 's}$  are a set of approximately Gaussian
weights. A gap is considered  significant if its value is greater than
2.25  \citep{Wainer76}.  The  presence  of big  gaps  in the  velocity
distribution indicates  that we are  not sampling a  single structure.
Fig.  \ref{fig:histoz}  (a) shows  the  redshift  histogram for  the
entire spectroscopic sample where  we have marked the main kinematical
structures, defined as those having  3~or more members, found from the
ROSTAT gap analysis.

The  dominating kinematical  structure  labelled D,  at redshift  $\sim
0.22$,  is (kinematically)  centered on  the brightest  galaxy  of the
cluster A1942 (\#067 in Table~A.1; z = 0.225). It neighbours structures C
and  E which,  given the  relatively small  redshifts  distances ($\rm
\Delta z  \sim 0.012$),  may, together with  structure D, belong  to a
same superstructure.   A new gap  analysis using this  subsample finds
the  same  gaps  as  the  previous  analysis,  indicating  that  these
structures are indeed kinematically distinct from each other.

Using the  sample of 44 redshifts  belonging to structure  D, which we
may identify  with the cluster A1942,  we may obtain  the cluster mean
recessional velocity, $\overline{cz} = 67493_{-249}^{+226}~\kms$, which
corresponds to  the redshift $z_{A1942} = z_D  = 0.22513$ \footnote{In
this paper  means and dispersions  are given as  biweighted estimates,
see \citet{Beers90}.   Error bars are 90\%  confidence intervals}. For
comparison, the  recessional velocity of  the brightest galaxy  of the
cluster  is $67402~\kms$. The  cluster velocity  dispersion corrected
following   \citet{Danese80}   is    found   to   be   $\sigma_{corr}=
908_{-139}^{+147}~\kms$.   We  notice  that  all the  normality  tests
included in the ROSTAT package fail to reject the null hypothesis of a
Gaussian  distribution  for  this  sample. The  neighbours  kinematical
structures  $C$ and  $E$ are  located at  $z_C =  0.20938$ and  $z_E =
0.23588$.

\begin{figure}[htb] 
\centering \includegraphics[width=\columnwidth]{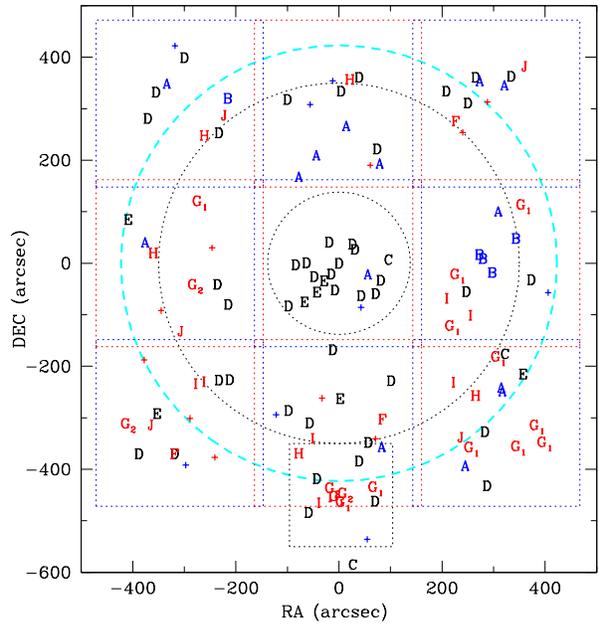}
\caption[]{The  projected  distribution  of the  spectroscopic  sample
galaxies (Table~A.1). Positions are  given as offsets from the brightest
A1942 galaxy,  (\#067 of Table~A.1). Galaxies are labelled  according to
the   kinematical   structures    they   belong,   given   in   Fig.
\ref{fig:histoz}  (a),  whereas  the  \textsl{crosses}  correspond  to
galaxies not assigned  to any of them.  The  \textsl{dotted lines} are
the boundaries of the EFOSC fields (Section \ref{sec:ObservationsReduction}), 
with the  most southern one centered at  the position of
the ``dark clump" suggested  by Erben \etal (2000). The \textsl{dashed
circle}  of  radius   $r_{200}  =  1.53\,h_{70}^{-1}~\rm{Mpc}$  ($\sim
420~\rm{arcsec}$)  show the  limits  of the  virialized cluster.   The
\textsl{dotted circles} have  radii $0.5\,h_{70}^{-1} \rm{Mpc}$ ($\sim
138~\rm{arcsec}$    )   and   $1.26\,h_{70}^{-1}    \rm{Mpc}$   ($\sim
350~\rm{arcsec}$,  roughly  corresponding   to  the  region  of  X-ray
emission).}
\label{fig:mapa_grupos}
\end{figure}

Fig. \ref{fig:mapa_grupos}  shows the projected  distribution of the
galaxies  corresponding   to  the  kinematical   structures  discussed
above. Notice that galaxies belonging to structures C, D and E seem to
populate the same regions  in projection, corroborating the suggestion
that they are part of  a bigger superstructure. The plot also suggests
that the main structure~D, which  is the cluster A1942 itself, extends
into a  large region, maybe  even larger than  the one covered  by our
observations.

\begin{figure}[htb]
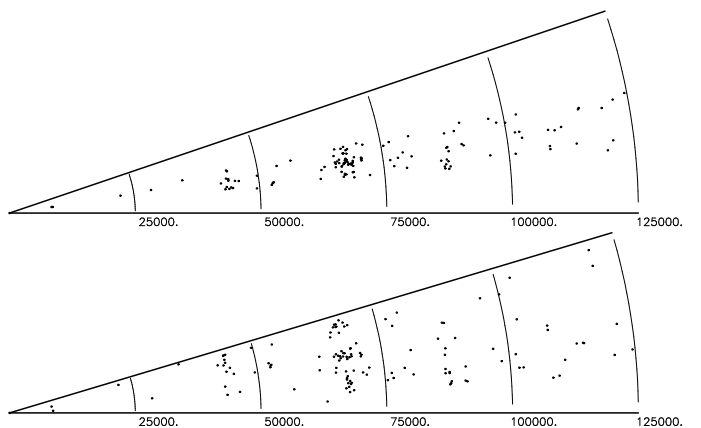

   \centering
   \includegraphics[height=\columnwidth,angle=-90]{alp1942.ps}
   \includegraphics[height=\columnwidth,angle=-90]{del1942.ps}
\caption[]{Wedge velocity  diagram in right  ascension (\textit{top}),
and declination  (\textit{bottom}) for the galaxies  in the Abell~1942
field.}
\label{fig:wedge}
\end{figure}

The wedge diagrams of galaxies  in right ascension and declination are
displayed in  Fig. \ref{fig:wedge}.  Both  show the cluster  and all
galaxies  collected in  a  square of  14.5~arcmin  around the  cluster
centre.   The  main  fore  and  background  concentrations  of  Fig.
\ref{fig:histoz}(a),   respectively   $A$   and   $G$   of   Fig.
\ref{fig:histoz}(a) are clearly seen in this figure.

\begin{figure}[htb]
\centering \includegraphics[width=\columnwidth]{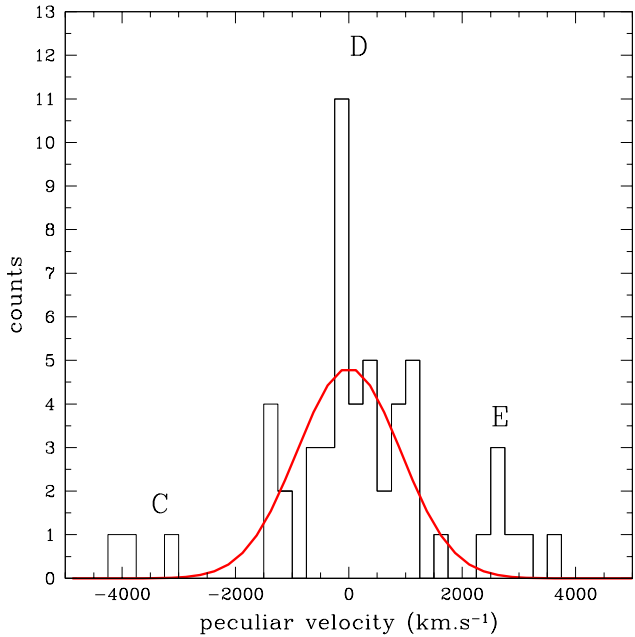}
\caption[]{Peculiar velocity  distribution in the  interval comprising
structures  C, D  and  E.  The  continuous  curve is  a Gaussian  with
dispersion  equal to  that  of structure  D (section  \ref{kinematical
structures}).}
\label{fig:histov_CDE}
\end{figure}

Fig. \ref{fig:histov_CDE} shows the peculiar velocity distribution, $v_{pec}
= c(z - z_D)/(1+z_D)$ of this sample. There are 3 galaxies belonging to the
``E'' group which are at the position corresponding to the X-ray substructure
at $1.7~\rm{arcmin}$ towards the southeast from the cluster centre (see
Section \ref{x-ray}). In any case, we cannot conclude of a heating process of
A1942 by a merging background group.

\subsection{The dark clump} \label{DC}

A massive dark  clump (hereafter DC), with total  mass equivalent to a
rich cluster  of galaxies and  situated about 7~arcmin south  to A1942
centre, was first suggested by \citet{Erben00} based on a weak lensing
signal detection  made on high  quality images obtained with  the CFHT
telescope. However,  a more detailed study  by \citet{Linden06}, using
HST quality observations, confirmed  the weak lensing signal detection
but with much less significance.  This finding, together with the fact
that Chandra observations of the  field did not showed any significant
extended X-ray emission in the DC direction (Section \ref{x-ray}), led
the authors  to scale down the total  mass of the cluster  by a factor
of, at  least, two as  compared to the original  value estimated by
\citet{Erben00}, making unlikely the  hypothesis of a dark matter halo
of the size of a galaxy  cluster. However, as the authors have pointed
out, there is  an noticeable excess of galaxies  in this region, which
is also apparent from the isopleths of the projected density of galaxies 
displayed in Fig.~\ref{fig:iso}. Note that the projected  density 
distribution nearby the  DC region is  elongated towards the same  direction
shown in the mass density maps  produced  by \citet{Linden06} (see  figure~9
of their paper), although being  much more extended than  it was found
there (the mass density enhancements found by \citet{Linden06} are mostly
concentrated inside the dashed circle delimiting the DC region).

We have spectroscopic data for 15 galaxies in the DC direction, most of them
in the background relative to A1942 , as can be seen in Fig.
\ref{fig:mapa_grupos}. The redshift distribution of these galaxies is shown in
panel (b) of Fig. \ref{fig:histoz}. As it can be seen by comparing with
panel (a), the DC redshift distribution may be decomposed into the same
kinematical groups as for the rest of the field. Hence, the spectroscopic data
indicate that, considering only galaxies in the cluster background, $z >
z_{D}$, there are two main kinematical structures, namely $G$ and $I$, that
may be contributing for the weak-lensing signal associated to the DC. However,
as it can be seen in Fig. \ref{fig:mapa_grupos}, these 2 structures are much
more spatially extended than the DC region, and tend to populate the entire
southern part of field, as, in fact, most of the background galaxies we
measured.

Additional clues on possible matter condensations that may be contributing to
the weak-lens signal in the direction of the DC can be obtained by looking at
the distribution of photometric redshifts. As part of an ongoing project of
estimating photometric redshifts for all galaxies brighter then $r=21$ in the
SDSS/DR6 using a procedure called Locally Weighted Regression briefly
described in the Appendix \citep[see also][]{Atkeson97,boris,abdalla}, we
have produced our own estimates of photometric redshifts for the region of
Fig.~\ref{fig:map}.
 
\begin{figure}[htb]
\centering
\includegraphics[width=\columnwidth]{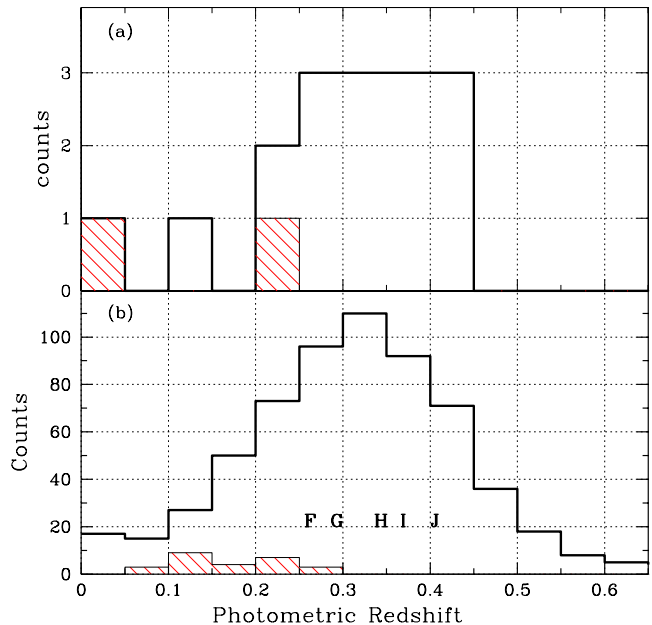}
\caption[]{The distribution of photometric redshifts  for galaxies  
projected: {\sl (a)} in a 120 arcsec radius circle enclosing the DC region;
{\sl (b)} in its complementary region within the $18 \times 18$~arcsec field 
of Fig. \ref{fig:iso}. Shaded histograms corresponds to the brighter sample 
of galaxies $15^{\rm mag} < m_{r} < 18^{\rm mag}$  and thick lines histogram
to the fainter sample $18^{\rm mag} < m_{r} < 21^{\rm mag}$.  Letters 
identify some of the kinematical structures found in 
Fig. \ref{fig:histoz}(a).}
\label{fig:hzphot}
\end{figure}

Fig. \ref{fig:hzphot} show the distribution of photometric redshifts
obtained with the LWR technique for galaxies belonging both to the DC region and
to its complementary region within the $18 \times 18$~arcsec field of Fig.
\ref{fig:iso}. Besides the fact that the brightest galaxies tend to avoid the DC
region (shaded histograms), for the fainter galaxies ($18^{\rm{mag}} < m_{r} <
21^{\rm{mag}}$) it can be seen that, whereas the redshift distribution of the
complementary DC field peaks for redshifts $\geq 0.3$ (bottom panel), within
the DC region it extends almost uniformly in the interval $0.3 < z_{LWR} <
0.45$.

\begin{figure}[htb]
\centering
\includegraphics[width=0.8\columnwidth]{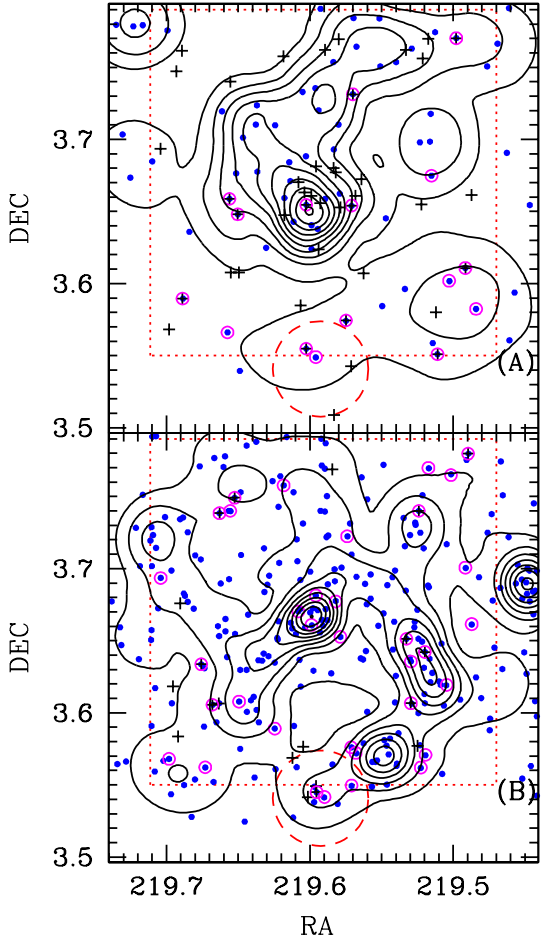}
\caption[]{Contours of the surface probability density of 
galaxies for 2 subsamples of faint galaxies selected accordingly to their 
photometric redshift. {\sl Upper panel}: $0.2 < z_{LWR} < 0.25$ ; 
{\sl lower panel}: $0.3 < z_{LWR} < 0.45$.  {\sl Filled circles} 
show the positions of galaxies of the selected samples;  those galaxies 
having a measured spectroscopic redshift are denoted by a large 
{\sl open circle}. Galaxies from the spectroscopic sample having $z_{\rm spec}$
in the same interval defining the sample, are denoted by {\sl crosses}.
Contours  are equally spaced by $\Delta_{\Sigma} = 0.12$ with the highest 
level at  $\Sigma_{Prob}^{max} = 5 \times 10^{-6}\,$arcsec$^{-2}$.
}
\label{fig:iso_DC}
\end{figure}

In order to see if the differences between the redshifts distributions have
some spatial counterparts, we show in Fig. \ref{fig:iso_DC} the surface
density of galaxies normalized by their total number\footnote{Since we are
using a continuously differentiable kernel for the density estimations, this
quantity is in fact a probability density.}, for two subsamples of faint
galaxies, selected accordingly to their photometric redshifts: galaxies with
photometric redshifts in the interval $0.2 < z_{LWR} < 0.25$ (panel A), which
in principle should have a greater probability to belong to cluster A1942; and
galaxies with $0.3 < z_{LWR} < 0.45$ (panel B) which constitute the major
fraction of galaxies projected in the DC region. Comparing these 2 panels, it
appears that galaxies with higher redshifts seem to have a slightly higher
probability to be found nearby the DC direction than galaxies in the low
redshift interval. This suggests that the density excess seen by
\citet{Erben00} should be due to line-of-sight background structures. Note
that galaxies at this redshift are too faint to be included in our
spectroscopic sample. Nevertheless, this is a hint of another structure that
may be contributing to the weak-lensing signal. In conclusion, our analysis
suggests that the DC weak-lensing signal, if not a mere statistical
fluctuation \citep{Linden06}, should probably be produced by several
structures along the line-of-sight.

\section{Summary and conclusion}

We have investigated the cluster of galaxies Abell~1942. More than a hundred
new spectroscopic redshifts were measured in a $14 \times 14$ region around
its centre. Together with X-ray archive data from Chandra, and photometric
data from SDSS, we were able to achieve the dynamical and kinematical analysis
of this cluster for the first time.

-- We found that about half of the observed galaxies are kinematical members
of the cluster. We have also found some kinematical evidence for the presence
of nearby groups of galaxies whose spatial counterparts however have not been
confirmed. The cluster is situated at redshift 0.22513.

-- Our analysis indicates that inside a radius of $\sim 1.7 h_{70}^{-1}\,$Mpc
($\sim 7\,$arcmin) the cluster galaxy distribution is well relaxed without any
remarkable feature and with a mean velocity dispersion of $\sigma=
908^{+147}_{-139}\kms$.

-- We have analyzed archival Chandra data and derived a mean temperature $kT =
5.5 \pm 0.5\,$keV and metal abundance $Z = 0.33 \pm 0.15 Z_{\odot}$. The
velocity dispersion corresponding to this temperature using the
$T_{X}$--$\sigma$ scaling relation is in good agreement with the measured
galaxies velocities. The X-ray emission traces fairly well the galaxy
distribution. 

-- We derive dynamical mass estimates of the cluster, assuming hydrostatic
equilibrium of the (isothermal) intracluster X-ray emitting gas. At a radius
equivalent to $r_{500}$ we obtained $M_{\rm dyn}(< r_{500}) \sim 5 \times
10^{14} M_{\odot}$. An estimate of the dynamical mass using the gravitational
arc found at about $ \sim 8.2$ arcsec from the cluster centre showed it to be
consistent with that derived from the hydrostatic equilibrium hypothesis.

-- We do not confirm the mass concentration 7~arcmin south from the cluster
centre from our dynamical and X-ray analysis. However, we do see a concentration
of background galaxies towards these regions which may be at the origin of the
weak lensing signal detected before.

\acknowledgements{We are grateful to an anonymous referee for valuable comments
that helped improve the paper. We thank the ESO astronomers Ivo Saviane and Gaspare Lo
Curto and staff for their assistance during the observations. HVC, LSJ and GBLN
thank the financial support provided by FAPESP and CNPq. DP acknowledges the
France-Brazil PICS-1080 and IAG/USP for its hospitality. GBLN thanks support
from the CAPES/COFECUB French-Brazilian cooperation program.}

\appendix
\section[]{}

Locally Weighted Regression (LWR) is a statistical method of a class known as memory-based
learning. In this type of method, a training set remains stored until a result, an answer
from a query, is obtained. LWR establishes a linear relationship between photometric feature
vectors (magnitudes or colours and/or additional photometric parameters) and redshifts which
is local because redshift estimation, at a given point in feature space, weighs more heavily
the data points in the neighbourhood of this point than those more distant.
For this work, we have used SDSS magnitudes (u, g, r, i, z) for the feature vectors. The training
set contains feature vectors and spectroscopic redshifts for all objects and, from these values,
we build a redshift estimator which is applied to the galaxies in the A1942 field.

For the training set, we have constructed a sample with 40951 unique galaxy spectroscopic redshifts,
using a similar procedure described in \citet{Oyaizu08}. We selected randomly 20000 redshifts from
SDSS-DR6 spectroscopic sample \citep{Adelman08}, with confidence level $z_{conf} > 0.9$. We also
included the following unique galaxy redshifts from other surveys: 290 from CFRS (Canada-France
Redshift Survey, \cite{Lilly95}), with $Class > 1$; 229 from DEEP (Deep Extragalactic Evolutionary
Probe, \cite{Davis01}) with $q_z =$ A or B; 6,245 from DEEP2 \citep{Weiner05} with
$z_{quality} \geq 3$; 414 from CNOC2 (Canadian Network for Observational Cosmology Field Galaxy
Survey, \cite{Yee00}); 13,257 from 2SLAQ (2dF-SDSS LRG and QSO Survey, \cite{Cannon06}) with
$z_{op} \geq 3$; and 416 from TKRS (Team Keck Redshift Survey, \cite{Wirth04}) with $z_{quality} > 1$. 

We have tested the method by comparing the photometric redshifts with the measured spectroscopic
ones (described in Section \ref{sec:ObservationsReduction}). These results are quoted in
Table~A.1. Using a sigma-clipping procedure in order to exclude catastrophic outliers
($\sim 5\%$ of the sample) we found a rms dispersion of 0.097 between $z_{phot}$ and $z_{spec}$. 

\makeatletter\if@referee\renewcommand\baselinestretch{1.0}\fi\makeatother
\begin{table*}[htb]
\caption[]{Positions, photometric data, photometric and spectroscopic redshifts
for galaxies of Abell~1942.}
\rotatebox{90}{
\small
\begin{tabular}{ccccccccccccl}
\hline
\hline
{\bf Gal.}&{\bf R.A.}&{\bf Decl}& {\bf $u$}&{\bf $g$}&{\bf $r$}&{\bf $i$}&{\bf $z$}&{\bf $z_{phot}$}&{error}&{\bf $z_{spec}$}&{error}&{notes} \\
      & (2000)      &     (2000) &         &         &         &         &         & (LWR)  &       &       &                  &               \\
\hline
   1 & 14 37 54.75 & +03 39 15.7 & 22.301 & 20.283 & 19.315 & 18.907 & 18.610 &  0.088 &0.052 & 0.16488 & 0.00032 & em: {$H\beta$},2OIII \\
   2 & 14 37 55.18 & +03 34 22.7 & 22.875 & 22.012 & 20.747 & 20.191 & 19.755 &  0.459 &0.075 & 0.29303 & 0.00014 & R: 4.85 \\
   3 & 14 37 56.19 & +03 34 54.9 & 25.629 & 21.395 & 19.982 & 19.437 & 18.886 &  0.221 &0.047 & 0.29046 & 0.00018 & R: 5.04 \\
   4 & 14 37 56.94 & +03 39 41.9 & 22.573 & 21.417 & 19.943 & 19.406 & 19.069 &  0.350 &0.057 & 0.22980 & 0.00028 & R: 3.02 \\
   5 & 14 37 57.52 & +03 46 50.9 & 23.519 & 22.251 & 20.674 & 20.148 & 19.854 &  0.339 &0.072 & 0.39893 & 0.00012 & em: OII \\
   6 & 14 37 57.98 & +03 42 03.4 & 20.923 & 20.545 & 19.734 & 19.428 & 19.128 &  0.415 &0.057 & 0.29414 & 0.00010 & em: OII \\
   7 & 14 37 58.00 & +03 36 38.2 & 22.116 & 20.677 & 19.533 & 19.059 & 18.812 &  0.249 &0.077 & 0.24038 & 0.00018 & R: 3.04 \\
   8 & 14 37 58.58 & +03 34 14.6 & 21.339 & 20.364 & 19.499 & 19.272 & 19.138 &  0.269 &0.048 & 0.29027 & 0.00032 & R: 3.12 very weak \\
   9 & 14 37 58.93 & +03 41 01.5 & 20.440 & 19.154 & 18.370 & 17.979 & 17.697 &  0.161 &0.038 & 0.17489 & 0.00015 & R: 6.12 \\  
 10 & 14 37 59.55 & +03 46 16.2 & 22.249 & 20.199 & 18.866 & 18.369 & 18.097 &  0.242 &0.039 & 0.22297 & 0.00012 & R: 8.47 \\
 11 & 14 38 00.38 & +03 37 17.2 & 23.322 & 21.762 & 21.536 & 21.339 & 21.695 &            &          & 0.20759 & 0.00016 & R: 3.37 \\
 12 & 14 38 00.42 & +03 45 56.6 & 25.051 & 20.510 & 19.558 & 19.074 & 18.817 &  0.325 &0.047 & 0.14665 & 0.00019 & R: 3.67 \\
 13 & 14 38 00.71 & +03 36 05.4 & 21.946 & 22.071 & 19.872 & 19.780 & 19.128 &  0.216 &0.042 & 0.14199 & 0.00019 & R: 3.02 \\
 14 & 14 38 00.83 & +03 36 12.5 & 21.851 & 19.601 & 18.663 & 18.245 & 17.891 &  0.096 &0.024 & 0.14403 & 0.00056 & em: OII,$H\beta$,2OIII \\
 15 & 14 38 01.17 & +03 37 08.4 & 21.853 & 19.828 & 18.205 & 17.659 & 17.301 &  0.303 &0.023 & 0.29125 & 0.00016 & R: 5.82 \\
 16 & 14 38 01.21 & +03 41 54.4 & 19.853 & 18.450 & 17.615 & 17.193 & 16.917 &  0.142 &0.027 & 0.14915 & 0.00016 & R: 4.90 em: OII $z$=0.15048 \\
      &           &             &        &        &             &             &             &            &          & 0.15006 &               & SDSS \\
 17 & 14 38 02.00 & +03 39 55.6 & 21.189 & 20.104 & 19.564 & 19.276 & 19.145 &  0.134 &0.042 & 0.17558 & 0.00013 & em: $H\beta$,2OIII \\
 18 & 14 38 02.48 & +03 45 26.4 & 22.480 & 20.698 & 22.549 & 19.353 & 19.569 &            &          & 0.26529 & 0.00038 & R: 3.09 \\
 19 & 14 38 02.66 & +03 33 01.2 & 23.727 & 20.840 & 19.613 & 19.158 & 18.871 &  0.231 &0.055 & 0.22390 & 0.00013 & R: 3.08 \\
 20 & 14 38 02.96 & +03 34 46.5 & 21.334 & 19.434 & 18.194 & 17.724 & 17.519 &  0.196 &0.023 & 0.22483 & 0.00008 & R: 10.67 \\
 21 & 14 38 03.21 & +03 40 22.3 & 20.371 & 19.098 & 18.264 & 17.662 & 17.352 &  0.156 &0.036 & 0.17596 & 0.00014 & R: 3.62 \\
      &           &             &        &        &             &             &             &            &          & 0.17548 & 0.00043 & em: OII,$H\beta$,2OIII \\
 22 & 14 38 03.62 & +03 46 07.7 & 20.601 & 19.248 & 18.463 & 18.034 & 17.765 &  0.138 &0.042 & 0.14806 & 0.00021 & R: 3.61 \\
 23 & 14 38 03.68 & +03 40 30.3 & 21.757 & 20.351 & 19.766 & 19.501 & 19.698 &  0.201 &0.046 & 0.17642 & 0.00021 & em: OII,$H\beta$,2OIII \\
 24 & 14 38 04.17 & +03 35 56.5 & 22.681 & 21.958 & 21.090 & 20.874 & 21.121 &  0.342 &0.098 & 0.33790 & 0.00014 & em: OII very weak \\ 
 25 & 14 38 04.20 & +03 46 14.7 & 22.609 & 21.378 & 20.237 & 19.729 & 19.337 &  0.385 &0.077 & 0.22394 & 0.00025 & R: 3.03 \\
 26 & 14 38 04.68 & +03 34 12.7 & 22.362 & 21.003 & 19.625 & 19.110 & 18.755 &  0.318 &0.059 & 0.29304 & 0.00014 & R: 4.68 \\
 27 & 14 38 04.80 & +03 38 31.5 & 23.483 & 22.528 & 20.851 & 20.272 & 19.915 &  0.436 &0.064 & 0.36106 & 0.00019 & R: 3.11 \\
 28 & 14 38 05.18 & +03 45 24.4 & 23.857 & 20.491 & 19.257 & 18.727 & 18.261 &  0.177 &0.038 & 0.22919 & 0.00019 & R: 4.13 \\
 29 & 14 38 05.37 & +03 39 17.9 & 22.576 & 21.189 & 19.989 & 19.554 & 19.158 &  0.293 &0.087 & 0.22980 & 0.00028 & R: 3.02 weak \\
 30 & 14 38 05.49 & +03 33 40.5 & 21.891 & 20.160 & 19.317 & 18.912 & 18.602 &  0.352 &0.065 & 0.14504 & 0.00022 & R: 4.75 \\
 31 & 14 38 05.82 & +03 44 26.5 & 21.756 & 20.838 & 19.485 & 18.924 & 18.676 &  0.372 &0.068 & 0.32107 & 0.00021 & R: 3.48 \\
 32 & 14 38 06.03 & +03 34 35.4 & 22.062 & 20.404 & 19.384 & 18.980 & 18.790 &  0.161 &0.067 & 0.40309 & 0.00027 & em: OII,$H\delta$,$H\beta$,2OIII \\
 33 & 14 38 06.45 & +03 39 47.6 & 22.943 & 21.288 & 19.747 & 19.217 & 18.856 &  0.339 &0.045 & 0.28833 & 0.00016 & R: 4.18 em: OII $z$= 0.28834 \\
 34 & 14 38 06.80 & +03 44 49.2 & 22.127 & 20.474 & 19.949 & 19.878 & 19.953 &            &          & 0.25692 & 0.00012 & em: $H\beta$ \\
 35 & 14 38 07.13 & +03 36 23.2 & 23.965 & 22.246 & 20.588 & 20.044 & 19.473 &  0.390 &0.045 & 0.37891 & 0.00021 & R: 4.18 \\
 36 & 14 38 07.13 & +03 38 07.4 & 24.399 & 22.260 & 20.468 & 20.043 & 19.768 &  0.406 &0.044 & 0.29209 & 0.00035 & R: 2.63 very weak \\
 37 & 14 38 07.86 & +03 39 04.8 & 21.160 & 19.789 & 18.216 & 17.651 & 17.293 &  0.302 &0.026 & 0.36109 & 0.00019 & R: 3.65 \\
 38 & 14 38 07.97 & +03 45 46.6 & 21.026 & 20.384 & 19.762 & 19.427 & 19.600 &  0.295 &0.047 & 0.22180 & 0.00014 & em: OII \\
 39 & 14 38 15.12 & +03 36 25.3 & 21.255 & 20.318 & 19.765 & 19.584 & 19.227 &  0.181 &0.043 & 0.23002 & 0.00013 & em: OII \\
 40 & 14 38 15.42 & +03 40 20.9 & 23.061 & 21.657 & 20.284 & 19.837 & 19.607 &  0.265 &0.080 & 0.20869 & 0.00018 & R: 3.04 very weak \\
 41 & 14 38 16.24 & +03 35 10.3 & 21.536 & 20.649 & 19.774 & 19.324 & 18.969 &  0.270 &0.051 & 0.25485 & 0.00020 & R: 3.09 very weak \\
 42 & 14 38 16.41 & +03 34 17.4 & 22.596 & 21.341 & 20.664 & 20.383 & 20.303 &  0.335 &0.078 & 0.15343 & 0.00026 & R: 3.01 \\
\hline
\end{tabular}}
\end{table*}
\begin{table*}[htb]
\rotatebox{90}{
\small
\begin{tabular}{ccccccccccccl}
\hline
\hline
{\bf Gal.}&{\bf R.A.}&{\bf Decl}& {\bf $u$}&{\bf $g$}&{\bf $r$}&{\bf $i$}&{\bf $z$}&{\bf $z_{phot}$}&{\bf $error$}&{\bf $z_{spec}$}&{error}&{notes} \\
          & (2000) & (2000) &          &         &         &         &         &  (LWR)  &                  &             &                  &               \\
\hline
 43 & 14 38 16.46 & +03 39 40.0 & 22.922 & 20.650 & 19.317 & 18.834 & 18.422 &  0.259 &0.043 & 0.22237 & 0.00011 & R: 7.46 \\
 44 & 14 38 16.58 & +03 43 26.7 & 20.507 & 19.752 & 19.339 & 18.988 & 19.037 &  0.163 &0.036 & 0.14155 & 0.00045 & em: $H\beta$,2OIII \\
 45 & 14 38 16.88 & +03 43 54.9 & 21.584 & 19.492 & 18.152 & 17.668 & 17.398 &  0.235 &0.022 & 0.21841 & 0.00013 & R: 6.82 \\
 46 & 14 38 17.09 & +03 39 14.5 & 24.485 & 20.968 & 19.739 & 19.263 & 19.063 &  0.228 &0.040 & 0.22257 & 0.00021 & R: 3.66 \\
 47 & 14 38 17.12 & +03 32 55.4 & 22.234 & 21.133 & 20.213 & 19.818 & 19.823 &  0.313 &0.062 & 0.29395 & 0.00021 & R: 3.90 \\
 48 & 14 38 17.15 & +03 34 32.0 & 22.210 & 21.069 & 19.866 & 19.263 & 18.889 &  0.407 &0.077 & 0.26918 & 0.00004 & em: $H\beta$ \\
 49 & 14 38 17.16 & +03 32 31.2 & 21.720 & 20.624 & 19.635 & 19.315 & 19.057 &  0.278 &0.057 & 0.22896 & 0.00028 & R: 3.28 \\
 50 & 14 38 17.83 & +03 43 23.2 & 23.431 & 21.609 & 20.348 & 19.810 & 19.302 &  0.352 &0.070 & 0.26016 & 0.00017 & R: 3.04 weak \\
 51 & 14 38 18.04 & +03 34 25.5 & 22.301 & 20.024 & 18.659 & 18.178 & 17.800 &  0.240 &0.029 & 0.22682 & 0.00009 & R: 10.73 \\
    &             &             &        &        &        &        &             &            &          & 0.22625 & 0.00013 & R: 6.62 \\
 52 & 14 38 18.12 & +03 39 52.4 & 22.875 & 20.492 & 19.459 & 19.136 & 18.893 &  0.123 &0.054 & 0.15097 & 0.00019 & R: 3.04 \\
 53 & 14 38 18.20 & +03 31 17.7 & 17.311 & 15.965 & 15.362 & 15.047 & 14.814 &  0.039 &0.016 & 0.02899 & 0.00020 & em: $H\beta$,$H\alpha$,SI \\
      &                     &                     &             &             &             &              &        &        &      & 0.02877 &         & SDSS \\
 54 & 14 38 18.93 & +03 38 22.5 & 23.251 & 22.207 & 21.814 & 21.666 & 21.927 &  0.681 &0.192 & 0.02816 & 0.00006 & em: $H\alpha$,NII \\
    &             &             &        &        &        &        &        &        &      & 0.02802 &         & SDSS \\
 55 & 14 38 18.97 & +03 39 09.8 & 22.256 & 20.970 & 19.718 & 19.227 & 18.844 &  0.321 &0.081 & 0.22833 & 0.00015 & R: 5.54 \\
 56 & 14 38 19.23 & +03 46 13.9 & 22.991 & 20.532 & 19.203 & 18.692 & 18.329 &  0.270 &0.038 & 0.22880 & 0.00013 & R: 6.42 \\
 57 & 14 38 19.27 & +03 33 49.3 & 20.389 & 18.982 & 17.694 & 17.256 & 16.943 &  0.235 &0.031 & 0.22719 & 0.00011 & R: 5.27 \\
    &             &             &        &        &        &        &        &        &      & 0.22625 & 0.00031 & R: 3.08 \\
 58 & 14 38 19.76 & +03 40 39.1 & 22.874 & 21.837 & 20.579 & 20.031 & 19.806 &  0.384 &0.073 & 0.22383 & 0.00019 & R: 3.05 \\
 59 & 14 38 20.02 & +03 30 27.9 & 21.294 & 20.017 & 19.247 & 18.872 & 18.719 &  0.157 &0.053 & 0.21106 &         & uncertain \\
 60 & 14 38 20.09 & +03 40 50.2 & 22.755 & 20.676 & 19.276 & 18.784 & 18.420 &  0.268 &0.061 & 0.22985 & 0.00033 & R: 3.09 \\
 61 & 14 38 20.41 & +03 46 10.4 & 22.100 & 21.369 & 20.538 & 20.746 & 20.110 &  0.296 &0.080 & 0.34347 & 0.00009 & em: OII \\
 62 & 14 38 20.96 & +03 44 39.8 & 19.853 & 18.004 & 16.937 & 16.511 & 16.140 &  0.138 &0.011 & 0.14667 & 0.00010 & R: 7.71 \\
    &             &             &        &        &        &        &        &        &      & 0.14688 &         & SDSS \\
 63 & 14 38 21.47 & +03 32 43.8 & 23.094 & 21.450 & 21.075 & 21.075 & 21.220 &  0.131 &0.105 & 0.30440 & 0.00012 & em: OII \\
 64 & 14 38 21.59 & +03 45 47.3 & 21.247 & 19.623 & 18.316 & 17.857 & 17.495 &  0.261 &0.028 & 0.22408 & 0.00010 & R: 8.12 \\
 65 & 14 38 21.63 & +03 35 51.1 & 22.705 & 21.938 & 21.277 & 20.982 & 20.928 &  0.359 &0.104 & 0.23596 & 0.00026 & R: 3.03 \\
 66 & 14 38 21.67 & +03 32 27.2 & 23.958 & 21.441 & 20.062 & 19.457 & 19.171 &  0.354 &0.037 & 0.29285 & 0.00021 & R: 4.65 \\
 67 & 14 38 21.85 & +03 40 12.9 & 19.720 & 17.557 & 16.226 & 15.698 & 15.331 &  0.168 &0.015 & 0.22483 & 0.00025 & R: 5.55 \\
    &             &             &        &        &        &        &        &        &      & 0.22475 &         & SDSS \\
 68 & 14 38 22.39 & +03 39 21.3 & 22.314 & 21.480 & 20.161 & 19.685 & 19.390 &  0.410 &0.095 & 0.22662 & 0.00044 & R: 3.01 weak \\
    &             &             &        &        &        &        &        &        &      & 0.22573 & 0.00029 & R: 3.01 weak \\
 69 & 14 38 22.55 & +03 32 36.8 & 23.520 & 21.309 & 20.153 & 19.598 & 19.307 &  0.260 &0.064 & 0.29351 & 0.00019 & R: 3.22 \\
 70 & 14 38 22.63 & +03 37 25.3 & 21.054 & 19.479 & 18.418 & 18.027 & 17.748 &  0.196 &0.036 & 0.21867 & 0.00021 & R: 3.02 \\
 71 & 14 38 22.64 & +03 46 07.4 & 20.394 & 19.406 & 18.722 & 18.370 & 18.211 &  0.195 &0.035 & 0.16666 & 0.00033 & R: 3.19 \\
   &              &             &        &        &        &        &        &        &      &         &         & em:OII,$H\beta$,2OIII $z$=0.16471 \\
 72 & 14 38 22.84 & +03 39 51.8 & 21.552 & 19.317 & 17.966 & 17.468 & 17.138 &  0.230 &0.017 & 0.22109 & 0.00012 & R: 8.56 \\
 73 & 14 38 22.98 & +03 32 39.3 & 22.065 & 21.421 & 20.172 & 19.629 & 19.316 &  0.434 &0.065 & 0.36555 & 0.00010 & R: 3.03 \\
 74 & 14 38 23.13 & +03 40 53.6 & 22.720 & 20.210 & 18.869 & 18.387 & 18.051 &  0.318 &0.031 & 0.23017 & 0.00024 & R: 3.62 \\
 75 & 14 38 23.15 & +03 32 53.9 & 20.972 & 19.839 & 19.074 & 18.676 & 18.454 &  0.211 &0.039 & 0.30181 & 0.00023 & R: 3.02 em: $z$= 0.30277 \\
    &             &             &        &        &        &        &        &        &      & 0.30318 & 0.00017 & R: 5.12 \\
 76 & 14 38 23.83 & +03 39 39.4 & 21.773 & 20.782 & 19.589 & 19.115 & 18.804 &  0.380 &0.079 & 0.23579 & 0.00013 & R: 6.69 \\
\hline
\end{tabular}}
\end{table*}
\begin{table*}[htb]
\rotatebox{90}{
\small
\begin{tabular}{ccccccccccccl}
\hline
\hline
{\bf Gal.}&{\bf R.A.}&{\bf Decl}& {\bf $u$}&{\bf $g$}&{\bf $r$}&{\bf $i$}&{\bf $z$}&{\bf $z_{phot}$}&{\bf $error$}&{\bf $z_{spec}$}&{error}&{notes} \\
          & $(2000)$ & $(2000)$ &          &         &         &         &         &  (LWR)  &                  &             &                  &               \\
\hline
 77 & 14 38 24.19 & +03 35 48.2 & 22.968 & 21.574 & 20.381 & 19.920 & 19.623 &  0.269 & 0.080 & 0.26374 & 0.00023 & uncertain \\
 78 & 14 38 24.43 & +03 32 26.8 & 21.937 & 20.926 & 20.256 & 20.192 & 19.883 &  0.136 & 0.065 & 0.36131 & 0.00025 & R: 3.04 \\
 79 & 14 38 24.73 & +03 33 14.6 & 22.114 & 20.126 & 18.846 & 18.340 & 17.986 &  0.220 & 0.039 & 0.22586 & 0.00008 & R: 10.74 \\
    &             &             &        &        &        &        &        &        &       & 0.22608 & 0.00010 & R: 8.15 \\
 80 & 14 38 24.73 & +03 39 17.3 & 22.317 & 20.532 & 19.251 & 18.779 & 18.488 &  0.231 & 0.053 & 0.23644 & 0.00013 & R: 6.53 \\
 81 & 14 38 24.80 & +03 43 42.9 & 20.938 & 18.910 & 17.809 & 17.331 & 16.989 &  0.165 & 0.015 & 0.14675 & 0.00012 & R: 6.85 \\
 82 & 14 38 25.04 & +03 39 47.5 & 20.779 & 19.689 & 18.901 & 18.609 & 18.472 &  0.193 & 0.040 & 0.21830 & 0.00038 & R: 3.15 \\
 83 & 14 38 25.27 & +03 34 33.1 & 22.222 & 21.431 & 20.966 & 20.740 & 20.422 &  0.272 & 0.098 & 0.37008 &         & em: OII \\
 84 & 14 38 25.60 & +03 45 20.7 & 19.274 & 18.234 & 17.928 & 17.680 & 17.566 &  0.053 & 0.027 & 0.07448 & 0.00011 & em: OII,$H\beta$,2OIII \\
 85 & 14 38 25.67 & +03 35 03.9 & 21.409 & 19.570 & 18.541 & 18.120 & 17.821 &  0.137 & 0.035 & 0.22499 & 0.00010 & R: 8.59 \\
    &             &             &        &        &        &        &        &        &       & 0.22524 & 0.00010 & R: 8.15 \\
 86 & 14 38 26.03 & +03 40 14.1 & 22.982 & 21.885 & 20.461 & 20.038 & 19.848 &  0.333 & 0.065 & 0.21985 & 0.00016 & R: 3.02 \\
 87 & 14 38 25.80 & +03 32 09.5 & 22.822 & 23.250 & 22.193 & 21.833 & 21.383 &  0.668 & 0.036 & 0.22622 & 0.00022 & R: 3.01 \\
 88 & 14 38 26.30 & +03 38 58.5 & 21.026 & 18.472 & 17.052 & 16.540 & 16.232 &  0.241 & 0.017 & 0.23593 & 0.00019 & R: 5.05 \\
    &             &             &        &        &        &        &        &        &       & 0.23549 &         & SDSS \\
 89 & 14 38 27.03 & +03 43 01.2 & 19.550 & 17.503 & 16.405 & 15.977 & 15.635 &  0.144 & 0.014 & 0.14532 & 0.00017 & R: 4.06 \\
    &             &             &        &        &        &        &        &        &       & 0.14621 &         & SDSS \\
 90 & 14 38 27.09 & +03 34 04.5 & 23.120 & 20.514 & 19.461 & 19.071 & 18.641 &  0.092 & 0.045 & 0.34219 & 0.00050 & weak \\
    &             &             &        &        &        &        &        &        &       & 0.33940 & 0.00050 & weak \\
 91 & 14 38 27.37 & +03 40 07.1 & 25.064 & 22.839 & 21.524 & 21.082 & 20.670 &  0.553 & 0.047 & 0.22561 & 0.00024 & R: 4.13 \\
 92 & 14 38 28.39 & +03 38 50.5 & 22.461 & 20.825 & 19.470 & 18.997 & 18.683 &  0.260 & 0.058 & 0.22131 & 0.00015 & R: 4.37 \\
 93 & 14 38 28.43 & +03 35 25.6 & 24.400 & 22.198 & 21.362 & 20.886 & 20.815 &  0.449 & 0.065 & 0.22557 & 0.00019 & R: 3.02 very weak \\
 94 & 14 38 28.54 & +03 45 30.8 & 23.794 & 23.036 & 20.741 & 20.093 & 19.901 &  0.403 & 0.044 & 0.23129 &         & uncertain \\
 95 & 14 38 30.00 & +03 35 18.4 & 22.482 & 21.345 & 19.838 & 19.322 & 18.905 &  0.354 & 0.062 & 0.09505 & 0.00023 & R: 3.10 \\
 96 & 14 38 35.98 & +03 36 26.6 & 21.731 & 20.405 & 19.267 & 18.867 & 18.612 &  0.320 & 0.069 & 0.21806 & 0.00014 & R: 6.40 \\
 97 & 14 38 36.21 & +03 38 52.6 & 23.617 & 20.411 & 19.112 & 18.621 & 18.335 &  0.217 & 0.032 & 0.22678 & 0.00013 & R: 5.99 \\
 98 & 14 38 36.30 & +03 45 32.9 & 23.684 & 21.893 & 21.179 & 21.024 & 20.955 &  0.216 & 0.041 & 0.17978 & 0.00045 & em: OII,$H\beta$,2OIII \\
 99 & 14 38 36.73 & +03 45 00.1 & 23.562 & 21.635 & 20.110 & 19.527 & 19.022 &  0.385 & 0.039 & 0.39874 & 0.00024 & R: 3.57 \\
100 & 14 38 37.40 & +03 36 26.6 & 21.741 & 20.215 & 18.877 & 18.408 & 18.055 &  0.275 & 0.046 & 0.22483 & 0.00021 & R: 4.01 \\
101 & 14 38 37.47 & +03 44 26.4 & 22.956 & 21.766 & 20.889 & 20.546 & 20.443 &  0.439 & 0.093 & 0.05753 & 0.22473 & 0.00021 R: 3.50 \\
102 & 14 38 37.59 & +03 39 31.9 & 21.825 & 19.670 & 18.394 & 17.880 & 17.530 &  0.231 & 0.024 & 0.02830 & 0.22427 & 0.00030 R: 4.12 \\
103 & 14 38 37.98 & +03 33 56.0 & 21.510 & 20.518 & 19.747 & 19.395 & 19.103 &  0.242 & 0.042 & 0.25180 & 0.00044 & R: 3.04 weak \\
104 & 14 38 38.33 & +03 40 43.4 & 22.905 & 19.760 & 18.606 & 18.110 & 17.679 &  0.059 & 0.028 & 0.62151 & 0.00024 & em: OII \\
105 & 14 38 39.29 & +03 44 20.9 & 21.654 & 21.366 & 20.396 & 20.120 & 19.559 &  0.300 & 0.078 & 0.33398 & 0.00015 & em: OII \\
106 & 14 38 39.42 & +03 36 22.1 & 21.334 & 21.310 & 20.142 & 19.736 & 19.408 &  0.451 & 0.055 & 0.38311 & 0.00022 & em: OII very weak \\
107 & 14 38 39.91 & +03 42 09.5 & 22.294 & 20.753 & 19.591 & 19.116 & 18.755 &  0.268 & 0.078 & 0.29289 & 0.00031 & R: 2.60 very weak \\
108 & 14 38 40.46 & +03 36 18.6 & 24.890 & 22.091 & 20.952 & 20.713 & 20.275 &  0.340 & 0.047 & 0.38371 & 0.00006 & em: OII \\
109 & 14 38 40.52 & +03 39 28.4 & 20.868 & 19.866 & 18.629 & 18.155 & 17.762 &  0.272 & 0.057 & 0.29957 & 0.00005 & em: OII,2OIII \\
110 & 14 38 41.17 & +03 35 13.1 & 22.309 & 21.852 & 21.153 & 20.822 & 20.640 &  0.538 & 0.098 & 0.32819 & 0.00011 & em: OII \\
111 & 14 38 41.72 & +03 33 41.4 & 22.291 & 22.062 & 20.516 & 20.044 & 19.596 &  0.381 & 0.063 & 0.18944 & 0.00018 & R: 3.02 \\
112 & 14 38 41.88 & +03 46 52.5 & 21.558 & 19.208 & 17.751 & 17.233 & 16.821 &  0.248 & 0.018 & 0.22659 & 0.00011 & R: 10.21 \\
113 & 14 38 42.19 & +03 34 15.8 &        &        &        &        &        &        &       & 0.25574 & 0.00026 & em: OII,$H\beta$,2OIII   ($\star$) \\
114 & 14 38 42.38 & +03 38 00.8 & 22.408 & 20.115 & 18.991 & 18.422 & 17.938 &  0.340 & 0.038 & 0.40693 &         & measured on H and K \\
115 & 14 38 43.13 & +03 47 15.7 & 18.994 & 17.334 & 16.472 & 16.080 & 15.769 &  0.094 & 0.021 & 0.11659 & 0.00011 & R: 9.50 \\
    &             &             &        &        &        &        &        &        &       & 0.11633 &         & SDSS \\
\hline
\end{tabular}}
\end{table*}
\begin{table*}[htb]
\rotatebox{90}{
\small
\begin{tabular}{ccccccccccccl}
\hline
\hline
{\bf Gal.}&{\bf R.A.}&{\bf Decl}& {\bf $u$}&{\bf $g$}&{\bf $r$}&{\bf $i$}&{\bf $z$}&{\bf $z_{phot}$}&{\bf $error$}&{\bf $z_{spec}$}&{error}&{notes} \\
    & (2000)    & (2000)    &          &         &         &         &         &     (LWR)      &             &                &             &              \\
\hline
116 & 14 38 43.20 & +03 34 02.6 & 20.200 & 19.075 & 17.956 & 17.445 & 17.172 &  0.210 &0.041 & 0.22769 & 0.00018 & R: 3.66 \\
117 & 14 38 44.34 & +03 46 02.7 & 22.847 & 21.024 & 19.962 & 19.585 & 19.747 &  0.023 &0.082 & 0.14677 & 0.00024 & R: 4.54 \\ 
118 & 14 38 44.90 & +03 38 41.6 & 20.823 & 19.396 & 18.763 & 18.474 & 18.399 &  0.081 &0.042 & 0.26911 & 0.00070 & uncertain \\
119 & 14 38 45.49 & +03 35 20.7 & 21.925 & 21.059 & 20.466 & 20.170 & 19.895 &  0.239 &0.075 & 0.23824 & 0.00030 & em: OII,$H\beta$,2OIII \\
120 & 14 38 45.59 & +03 45 45.1 & 23.102 & 21.756 & 20.684 & 20.139 & 20.187 &  0.674 &0.084 & 0.22347 & 0.00022 & R: 3.19 \\
121 & 14 38 45.88 & +03 40 33.4 & 21.710 & 20.690 & 19.764 & 19.369 & 18.973 &  0.271 &0.038 & 0.34203 & 0.00026 & R: 3.04 em: OII $z$= 0.34238 \\
122 & 14 38 46.23 & +03 34 59.1 & 25.929 & 22.260 & 20.808 & 20.204 & 19.685 &  0.574 &0.047 & 0.41526 & 0.00020 & R: 3.09 weak \\
123 & 14 38 46.60 & +03 44 53.2 & 25.893 & 22.189 & 20.838 & 20.277 & 19.455 &  0.613 &0.047 & 0.21933 & 0.00018 & R: 3.01 very weak \\
124 & 14 38 46.99 & +03 40 54.5 & 19.751 & 18.323 & 17.537 & 17.133 & 16.861 &  0.122 &0.027 & 0.14636 & 0.00021 & R: 3.09 \\
    &             &             &        &        &        &        &        &        &      & 0.14675 &         & SDSS \\
125 & 14 38 47.13 & +03 37 04.4 & 21.057 & 21.050 & 20.447 & 20.451 & 20.669 &  0.148 &0.107 & 0.32344 & 0.00012 & em: OII \\
126 & 14 38 47.78 & +03 34 02.7 & 24.408 & 21.495 & 20.163 & 19.775 & 19.359 &  0.311 &0.047 & 0.22545 & 0.00016 & R: 4.75 \\
127 & 14 38 49.14 & +03 41 37.7 & 22.913 & 21.242 & 19.542 & 18.985 & 18.646 &  0.354 &0.036 & 0.23440 & 0.00006 & em: $H\beta$ \\
128 & 14 38 49.16 & +03 34 58.6 & 22.195 & 21.919 & 21.093 & 20.786 & 20.297 &  0.490 &0.091 & 0.30402 & 0.00032 & R: 3.02 very weak \\
\hline
\end{tabular}}

Note to Table~A.1: \\
$\star$~ galaxy \# 113 is too faint: no photometric data found in SDSS database. \\

\end{table*}


\end{document}